\documentclass[reprint,prb,aps,floatfix,superscriptaddress]{revtex4-2}
\pdfoutput=1
\usepackage{amsmath}
\usepackage{amsfonts}
\usepackage{graphicx}
\usepackage{chemformula}
\usepackage{mathtools}
\usepackage{tabularx}
\usepackage{color}
\usepackage{bm}
\usepackage{dcolumn}
\usepackage{mathrsfs}

\renewcommand{\mathbf}[1]{\bm{#1}}

\begin{document}

\begin{abstract}
We present experimental and theoretical evidence of novel bound state formation 
in the low transverse field ordered phase of the quasi-one-dimensional 
Ising-like material CoNb2O6. High resolution single crystal inelastic neutron 
scattering measurements observe that small transverse fields lead to a breakup 
of the spectrum into three parts, each evolving very differently upon increasing 
field. This can be naturally understood starting from the excitations of the 
ordered phase of the transverse field Ising model, domain wall quasiparticles 
(solitons). Here, the transverse field and a staggered off-diagonal exchange 
create one-soliton hopping terms with opposite signs. We show that this leads to 
a rich spectrum and a special field, when the strengths of the off-diagonal 
exchange and transverse field match, at which solitons become localized; the 
highest field investigated is very close to this special regime. We solve this 
case analytically and find three two-soliton continua, along with three novel 
bound states. Perturbing away from this novel localized limit, we find very good 
qualitative agreement with the experimental data. We also present calculations 
using exact diagonalization of a recently refined Hamiltonian model for CoNb2O6 
and using diagonalization of the two-soliton subspace, both of which provide a 
quantitative agreement with the observed spectrum. The theoretical  models 
qualitatively and quantitatively capture a variety of non-trivial features in 
the observed spectrum, providing insight into the underlying physics of bound 
state formation.
\end{abstract}

\author{Leonie~Woodland}
\thanks{These authors contributed equally to this work}
\affiliation{Clarendon Laboratory, University of Oxford Physics Department, 
Parks Road, Oxford, OX1 3PU, UK} 
\author{Izabella~Lovas}
\thanks{These authors contributed equally to this work}
\affiliation{Kavli Institute for Theoretical Physics, University of California, 
Santa Barbara, 93106, California, USA}
\author{M. Telling} 
\affiliation{ISIS Facility, Rutherford Appleton Laboratory, Chilton, Didcot OX11 
0QX, UK}
\author{D. Prabhakaran} 
\affiliation{Clarendon Laboratory, University of Oxford Physics Department, 
Parks Road, Oxford, OX1 3PU, UK} 
\author{Leon~Balents}
\affiliation{Kavli Institute for Theoretical Physics, University of California, 
Santa Barbara, 93106, California, USA}
\affiliation{Canadian Institute for Advanced Research, Toronto, Ontario, Canada}
\author{Radu~Coldea}
\affiliation{Clarendon Laboratory, University of Oxford Physics Department, 
Parks Road, Oxford, OX1 3PU, UK} 

\title{Excitations of quantum Ising chain \ch{CoNb2O6} in low transverse field: 
quantitative description of bound states stabilized by off-diagonal exchange and 
applied field}
\date{\today}

\maketitle

\section{Introduction}

The transverse field Ising chain (TFIC) is an important model in condensed 
matter physics because it displays the key paradigms of both a continuous 
quantum phase transition from an ordered phase to a quantum paramagnetic phase 
as a function of field, as well as, in the ordered phase, of fractionalization 
of local spin flips into pairs of domain wall quasiparticles (solitons) 
\cite{Sachdev1999}. The pure TFIC model can be mapped to non-interacting 
fermions \cite{Pfeuty1970, Lieb1961}, which in the ordered phase represent these 
solitons. However, a variety of different additional subleading terms in the 
spin Hamiltonian, such as a longitudinal field \cite{McCoy1978} or an XY 
exchange \cite{Coldea2010} can stabilize two-soliton bound states. Here we 
explore a regime where novel bound states can be stabilized by the interplay of 
applied transverse field and off-diagonal exchange.

The material \ch{CoNb2O6} has been seen as a realization of TFIC physics for 
over a decade 
\cite{Coldea2010,Kinross2014xe,Morris2014ux,Liang2015,Amelin2020ov}. Among the 
key experimental observations is the qualitative change in the nature of 
quasiparticles from domain walls in the ordered phase to coherently propagating 
spin flips in the high-field paramagnetic phase. Moreover, a fine structure of 
bound states was observed just below the critical transverse field, consistent 
with predictions for a universal E8 spectrum expected in the presence of a 
perturbing longitudinal field, which in this case arises from mean-field effects 
of the three dimensional magnetic order \cite{Coldea2010,Amelin2020ov}. The 
crystal structure is orthorhombic (space group $Pbcn$), with \ch{Co^{2+}} ions 
with effective spin-$\frac{1}{2}$ arranged in zigzag chains running along the 
$c$-axis, with dominant nearest-neighbour ferromagnetic Ising exchange (see 
Fig.~1A). At the lowest temperatures, small three-dimensional interactions 
between chains stabilize a ground state with ferromagnetic ordering along the 
zigzag chains and with an antiferromagnetic pattern between chains 
\cite{Maartense1977ef, Scharf1979gi, Mitsuda1994, Heid1995bt, Weitzel2000ei}. 
While the dominant magnetic physics in \ch{CoNb2O6} can be captured by a TFIC 
Hamiltonian, additional terms in the Hamiltonian beyond the dominant Ising 
exchange are needed to explain various features of the spectrum 
\cite{Coldea2010, Fava2020}. In particular, a staggered off-diagonal exchange 
term was recently proposed on symmetry grounds and shown to reproduce well the 
zero-field spectrum using density matrix renormalization group numerics 
\cite{Fava2020}.

\begin{figure}
{\centering
\includegraphics[width=0.3\textwidth]{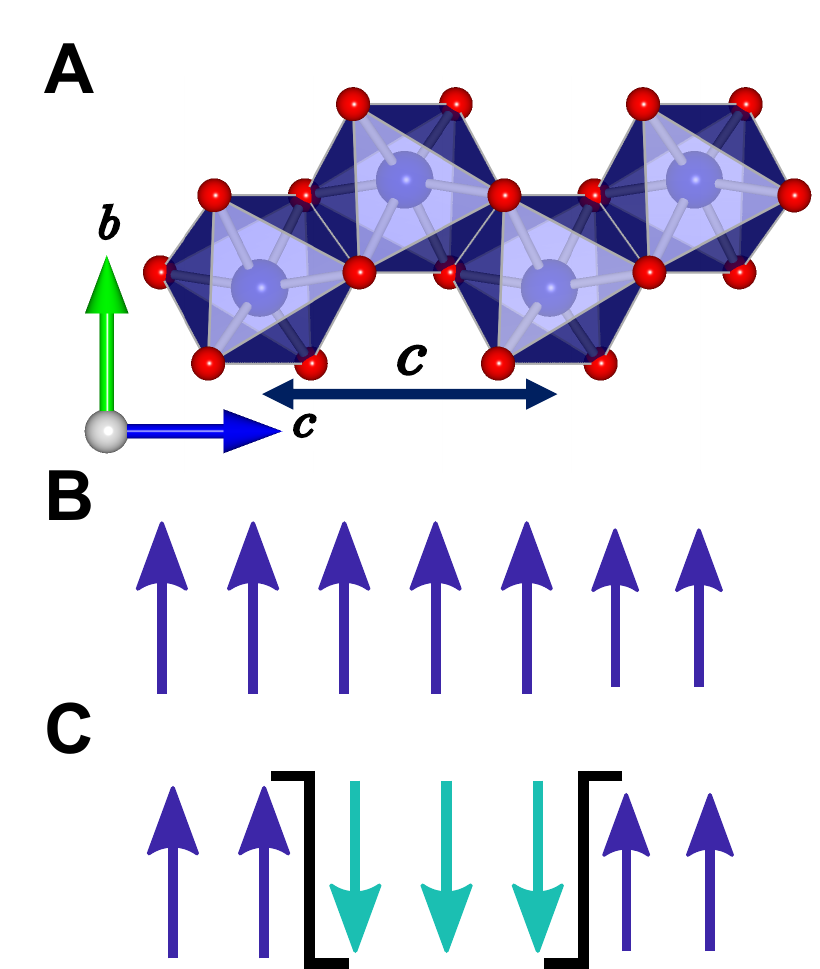}}
\caption{A) A single zigzag chain in \ch{CoNb2O6}. Blue spheres represent 
\ch{Co^{2+}} ions, red spheres \ch{O^{2-}} ions. The crystallographic unit cell 
length along the chain direction, $c$, is indicated. B) Ground state and C) 
excited state of a ferromagnetic Ising chain, obtained by reversing a spin 
cluster in the center. Only the boundaries (solid lines) cost energy, and they 
are the left and right solitons respectively. A spin-flip neutron scattering 
process creates such a pair of left and right solitons.}\label{F:zigzag}
\end{figure}

In this work, we present high resolution single crystal inelastic neutron 
scattering (INS) data as a function of low to intermediate transverse field in 
the ordered phase. This regime has also been explored by THz spectroscopy 
\cite{Morris2021}, which probe the zone-centre ($\mathbf{Q}=0$) excitations. The 
INS data reveal a rich evolution of the magnetic spectrum with increasing field: 
the spectrum splits into three parts with each part behaving very differently. 
The top two parts are sharp modes, with the top mode becoming progressively 
flatter and the middle one progressively more dispersive in field, while the 
lowest energy part is a continuum where the intensity moves from bottom to top 
upon increasing field. 

We seek to understand this rich behaviour in terms of a recently refined 
Hamiltonian model for \ch{CoNb2O6}, which proposed all relevant additional 
exchange terms down to 2\% of the Ising exchange \cite{Woodland2023}. We find 
that the experimental data agree very well with the results obtained using 
numerical exact diagonalization (ED) calculations for this full Hamiltonian, 
where interchain coupling effects are treated in a mean field approximation. To 
obtain a physical understanding of the spectrum, we start from a picture of 
soliton quasiparticles, which hop due to both the applied transverse field and 
the off-diagonal exchange. The competition between these hopping effects leads 
to soliton hopping terms that alternate along the two legs of the zigzag chain, 
resulting in two bands with dispersions that are tuned by the applied field. The 
relevance of a model with alternating hopping of solitons for explaining 
features in THz spectroscopy data obtained on \ch{CoNb2O6} was already mentioned 
in \cite{Morris2021}, where solitons were treated as non-interacting. Here we 
take into account fully the interactions between solitons as we find that this 
is crucial for a full understanding of the observed spectrum. A spin-flip 
neutron scattering process creates two-soliton excitations 
(Fig.~\ref{F:zigzag}C), which interact via hard-core repulsion and various 
nearest-neighbour interaction terms. We solve a minimal model in the two-soliton 
subspace and find three continua and up to three bound states depending on the 
values of the Hamiltonian parameters. To understand the character of these bound 
states, we first focus on the limit where solitons are localized due to the 
hopping term on alternate bonds being zero, a theoretical situation not 
previously explored. In this limit, novel bound states arise due to hard core 
repulsion. We then perturb away from this limit in first order perturbation 
theory, obtaining analytic expressions for the dispersions and intensities in 
INS, which give strong qualitative agreement with the data. The results indicate 
that this regime is indeed realized in \ch{CoNb2O6} at intermediate transverse 
field.

The rest of this paper is organized as follows: Sec. \ref{S:expdetails} provides 
details of the inelastic neutron scattering experiments while Sec. 
\ref{S:qualitativeexp} introduces the model Hamiltonian and provides a 
qualitative overview of the experimental results. In Sec. \ref{S:twosoliton}, we 
solve the model Hamiltonian in first order perturbation theory in the 
two-soliton subspace. In Sec. \ref{S:localized}, we provide a physical picture 
of the spectrum by starting from the limit where individual solitons are 
localized and perturbing around this limit. Sec. \ref{S:conclusion} contains our 
conclusions, and the Appendices give further technical details of the 
calculations.

\section{Experimental details}\label{S:expdetails}
Inelastic neutron scattering measurements of the magnetic excitations were 
performed on a large single crystal (6.76~g) of \ch{CoNb2O6} grown using a 
floating-zone technique \cite{PB} and already used in previous INS experiments 
\cite{Coldea2010}. The magnetic field was applied along the crystallographic 
$b$-direction, which is transverse to the local Ising axis of all the spins. The 
measurements were performed using the indirect geometry time-of-flight 
spectrometer OSIRIS at the ISIS facility. OSIRIS was operated with PG(002) 
analyzers to measure the inelastic scattering of neutrons with a fixed final 
energy of $E_f=1.82$~meV as a function of energy transfer and wavevector 
transfer in the horizontal $(h0l)$ scattering plane. Throughout this paper, we 
express the wavevector transfer in the inelastic neutron scattering experiments 
as $\mathbf{Q}=(2\pi h/a,0,2\pi l/c)$ where $(h,0,l)$ are expressed in 
reciprocal lattice units of the orthorhombic structural unit cell, with lattice 
parameters ${a=14.1337}$~\AA, $b=5.7019$~\AA ~ and $c=5.0382$~\AA ~ at 2.5~K 
\cite{Heid1995bt}. The sample was attached to the cold finger of a dilution 
refrigerator inside a vertical 7.5~T cryomagnet and measurements were taken at a 
temperature of 0.1~K. The average counting time at each field was around 
7~hours.

\begin{figure*}
\includegraphics[width=\textwidth]{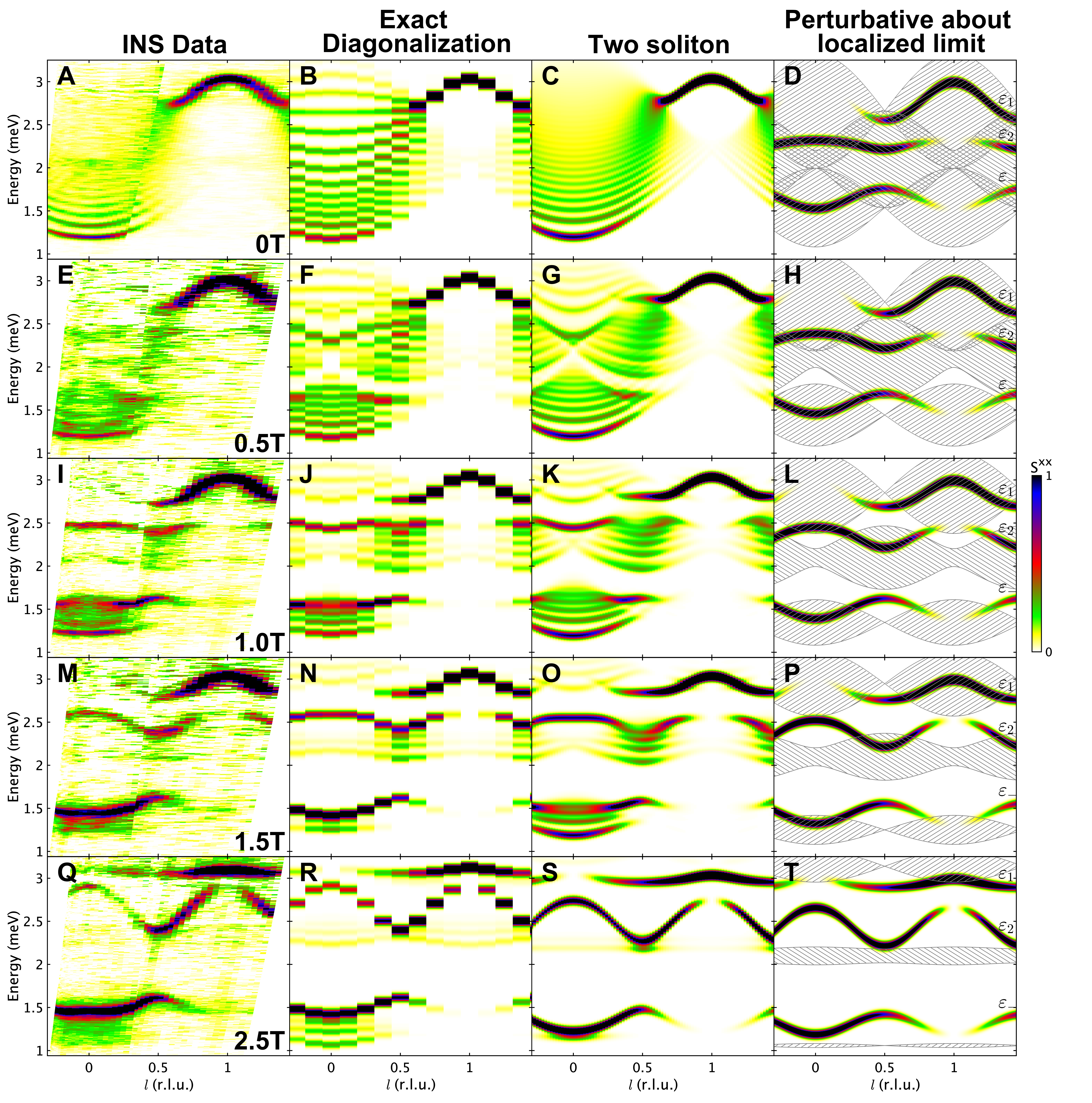}
\caption{ Evolution of the INS spectrum and calculated spectrum for the full 
Hamiltonian in (\ref{E:Hamiltonian}), as a function of applied magnetic field, 
increasing from top to bottom. Columns left to right: Inelastic neutron 
scattering data (A adapted from \cite{Coldea2010}); $S^{xx}$ as calculated by ED 
on 16 sites (8 unit cells) with periodic boundary conditions; $S^{xx}$ as 
calculated by diagonalization of the two-soliton subspace (see Sec. 
\ref{S:twosolitonsolution}) on 100 sites (50 unit cells) with periodic boundary 
conditions; $S^{xx}$ as calculated analytically by perturbing about the 
localized limit for $\lambda_{\text{MF}}$=0 (see Sec. \ref{S:localized}). In the 
right-most column, the hatched patches represent the two-soliton continua; in 
this approximation, the continua have no scattering intensity except where the 
bound states overlap them and create a resonance. In each panel, color indicates 
$S^{xx}$, as defined in (\ref{E:Sxx}) and further normalized by the total number 
of sites, on a linear scale indicated by the colorbar. The calculations have 
been convolved with a Gaussian of FWHM $0.067$~meV to mimic the estimated 
experimental resolution effects and calculated intensities are shown in absolute 
units of meV$^{-1}$.
The intensities for the data in panels E, I, M, Q, have been multiplied by a 
common scale factor to bring them visually into agreement with the corresponding 
calculations in column 2. The data in A come from a different experiment so a 
separate scale factor was used for those intensities to bring them into 
agreement with those in panel B.
}\label{F:expcomparison}
\end{figure*}

For each field, two sample orientations were measured ($c$-axis oriented in the 
scattering plane at 25$^{\circ}$ and 60$^{\circ}$ with respect to the incident 
beam direction). Throughout this paper, the data panels presented are a 
combination of data from these two orientations, with the wavevector projected 
along the chain direction $l$ as the physics considered is one-dimensional. The 
two orientations were chosen such that the projected $l$ values covered a large 
part of the Brillouin zone along the chain direction. The INS data at one of the 
measured fields (2.5~T, in Fig.~\ref{F:expcomparison}Q) was briefly reported in 
\cite{Fava2020}.  

The data shown have had an estimate of the non-magnetic background subtracted 
off, and have then been divided by the squared isotropic \ch{Co^{2+}} magnetic 
form factor $f^2(\mathbf{Q})$ and by the neutron polarization factor. The latter 
was calculated under the assumption that all inelastic scattering is in the 
polarizations perpendicular to the Ising ($z$) axes and that the dynamical 
structure factor satisfies 
$S^{xx}(\mathbf{Q},\omega)=S^{yy}(\mathbf{Q},\omega)$, an approximation which is 
found to be valid to a large extent for the model Hamiltonian 
(\ref{E:Hamiltonian}) in the low transverse field regime. Here,
\begin{equation}\label{E:Sxx}
S^{xx}(\mathbf{Q},\omega)=\sum_{\lambda_f}|\langle\lambda_f|S^{x}(\mathbf{Q})|
\text{GS}\rangle|^2\delta(E_{\lambda_f}-\hbar\omega),
\end{equation}
where the sum extends over all excited states $|\lambda_f\rangle$ of energy 
$E_{\lambda_f}$ relative to the ground state $|\text{GS}\rangle$
and where $S^x(\mathbf{Q})=\sum_j\exp(i\mathbf{Q}\cdot\mathbf{r}_j)S^x_j$,  with 
$j$ running over all sites. Under the above assumption, the wavevector 
dependence of the neutron polarization factor is
\begin{equation}\label{E:polarization}
\mathscr{P}(\mathbf{Q})=1+\frac{\left(\frac{2\pi 
h}{a}\right)^2\sin^2{\gamma}+\left(\frac{2\pi 
l}{c}\right)^2\cos^2{\gamma}}{\mathbf{Q}^2}.
\end{equation}
Dividing the raw inelastic neutron scattering intensities by 
$\mathscr{P}(\mathbf{Q})f^2(\mathbf{Q})$ then gives $S^{xx}(\mathbf{Q},\omega)$ 
up to an overall scale factor. Eq. (\ref{E:polarization}) is appropriate for the 
experimentally observed zero-field magnetic structure of \ch{CoNb2O6} and takes 
into account the two different chains per crystallographic unit cell with Ising 
directions at an angle of $\pm\gamma$ to the $c$-direction in the $ac$-plane. We 
have taken $\gamma$ to be 30$^{\circ}$\cite{Heid1995bt}.

\section{Evolution of the magnetic excitations with applied 
field}\label{S:qualitativeexp}
In this section, we first introduce the model Hamiltonian and relate this to the 
zero field spectrum, introducing the concept of two-soliton states. In applied 
field, the spectrum splits into three components with different evolution in 
field. We show in the following sections that this rich behaviour can be 
naturally understood in a picture of solitons with dispersions tuned by the 
transverse field.

\subsection{Model Hamiltonian}

We use the single-chain Hamiltonian model for \ch{CoNb2O6} recently refined in 
\cite{Woodland2023}.
It is convenient to write this in three parts:
\begin{equation}\label{E:Hamiltonian}
\mathcal{H}=\mathcal{H}_1 + \mathcal{H}_2 + \mathcal{H}_3
\end{equation}
where
\begin{align}
\mathcal{H}_1 = J &\sum_j \left[-S_j^zS_{j+1}^z 
-\lambda_S\left(S_j^xS_{j+1}^x+S_j^yS_{j+1}^y\right)\right.\nonumber\\
+&\left.(-1)^j\lambda_{yz}\left(S_j^yS_{j+1}^z+S_j^zS_{j+1}^y\right)\right]+
\sum_j 
h_yS_j^y , \label{E:Hmin}\\
\mathcal{H}_2 = J &\sum_j \left[-\lambda_A\left(S_j^xS_{j+1}^x - 
S_j^yS_{j+1}^y\right) \right.\nonumber\\
+ &\left.\lambda_{\text{AF}}S_j^zS_{j+2}^z 
+\lambda_{\text{AF}}^{xy}\left(S_j^xS_{j+2}^x+S_j^yS_{j+2}^y\right)\right] , 
\label{E:H2} \\
\text{and} \nonumber\\
\mathcal{H}_3 = J &\sum_j 2\lambda_{\text{MF}}\left(\langle S^y\rangle S^y_j - 
\langle S^z\rangle S^z_j\right),\label{E:H3}
\end{align}
and where $z$ is the Ising direction (in the crystallographic $ac$ plane, 
defined as the direction of the moments in zero applied field), $y$ is parallel 
to the crystallographic $b$-direction and $x$ completes a right-handed 
coordinate system. 

This Hamiltonian, with parameter values in Table~\ref{T:hamparams}, is a 
refinement of the minimal model proposed in Ref.~\cite{Fava2020} and was 
recently deduced using a simultaneous fit to the spectrum in zero field, large 
transverse field, and large near-longitudinal field \cite{Woodland2023}. One can 
regard $\mathcal{H}_1$ as the minimal Hamiltonian needed to \emph{qualitatively} 
reproduce all key features of the excitation spectrum, while the terms in 
$\mathcal{H}_2$ are sub-leading and are added in order to achieve 
\emph{quantitative} agreement. This applies to both the data in 
Ref.~\cite{Woodland2023} as well as data in low transverse field presented here. 
Finally, $\mathcal{H}_3$ captures the effects of the weak interchain 
interactions at a mean-field level. The given form with a constant 
$\lambda_{\text{MF}}>0$ applies throughout the field range explored here (0 to 
$2.5$~T~$\parallel b$) as the magnetic order pattern between chains does not 
change in this field range \cite{Wheeler2007} and as, for this magnetic 
structure, all the interchain interactions that have a net contribution to the 
mean field are Heisenberg-like \cite{Woodland2023}. 
The different signs in front of the $S_j^y$ and $S_j^z$ terms reflect the fact 
that the $S^y$-components of spins on neighbouring chains are parallel 
(polarized by the applied field $h_y$), whereas the $S^z$ components are 
(spontaneously) aligned antiparallel by the antiferromagnetic interchain 
interactions. 

The dominant term in the model Hamiltonian is the first term in $\mathcal{H}_1$, 
the nearest neighbour ferromagnetic Ising exchange.
The second term is a nearest neighbour ferromagnetic XY exchange term 
($\lambda_S$), which causes single spin flips to hop.
The third term is a staggered off-diagonal exchange term ($\lambda_{yz}$) which 
causes solitons to hop with a sign that alternates along the legs of the zigzag 
chain.
The transverse field term ($h_y=g_y\mu_BB_y$) flips single spins; for a fixed 
number of domain walls, this is equivalent to soliton hopping by one site. We 
will show that the competition between these two one-soliton hopping terms leads 
to a rich field-dependent spectrum.

The spectrum in zero field (Fig.~\ref{F:expcomparison}A data, 
\ref{F:expcomparison}B calculation) can be qualitatively understood in terms of 
the minimal Hamiltonian $\mathcal{H}_1+\mathcal{H}_3$. A spin-flip neutron 
scattering event creates a pair of solitons (domain walls) and for the pure 
Ising chain, the energy is independent of the separation between these solitons 
[see Fig.~\ref{F:zigzag}C]. In the presence of the staggered off-diagonal 
exchange ($\lambda_{yz}$), the solitons become dispersive \cite{Fava2020}, 
resulting in a continuum of scattering in energy-momentum space, covering a 
large energy extent near $l=0$.
A longitudinal interchain mean field ($\mathcal{H}_3$) acts as an effective 
linear potential confining the solitons into a series of bound states, as seen 
in Fig.~\ref{F:expcomparison}A near $l=0$. The sharp mode in the data near $l=1$ 
is a two-soliton kinetic bound state stabilized by the XY ($\lambda_S$) exchange 
\cite{Coldea2010}.

The terms in $\mathcal{H}_2$ do not change the qualitative content of the 
spectrum but are important for quantitative agreement with the experimental 
data. The first term in $\mathcal{H}_2$ is an antisymmetric diagonal nearest 
neighbour exchange ($\lambda_A$). The second and third terms are a next nearest 
neighbour antiferromagnetic XXZ exchange ($\lambda_{\text{AF}}$ and 
$\lambda_{\text{AF}}^{xy}$ respectively). The second term is needed to account 
for the energy of the kinetic bound state near $l=1$ \cite{Coldea2010}, while 
the first and third are needed to explain the details of the dispersions seen in 
very high field \cite{Woodland2023}. Fig.~\ref{F:expcomparison}B demonstrates 
that very good quantitative agreement has been achieved between the experimental 
zero-field data and exact diagonalization calculations using the full 
Hamiltonian (\ref{E:Hamiltonian}).

\begin{table}
\centering
\begin{tabular}{cd}
\hline
\hline
$J$& 2.48(2)\text{~meV}\\
$\lambda_S$& 0.251(6)\\
$\lambda_{yz}$ & 0.226(3)\\
$g_y$ & 3.32(2)\\
$\lambda_A$ & -0.021(1)\\
$\lambda_{AF}$ & 0.077(3)\\
$\lambda_{AF}^{xy}$ & 0.031(1)\\
$\lambda_{\text{MF}}$ & 0.0158(2)\\
\hline
\hline
\end{tabular}
\caption{Hamiltonian parameters used in (\ref{E:Hmin})-(\ref{E:H3}), from Ref. 
\cite{Woodland2023}}\label{T:hamparams}
\end{table}

\subsection{Spectrum in small to intermediate transverse field}

The key feature of the evolution of the spectrum as a function of field, shown 
in the first column of Fig.~\ref{F:expcomparison}, is the break-up of the 
observable spectrum into three parts, each evolving very differently upon 
increasing field. The top part, which evolves out of the zero-field high-energy 
kinetic bound state, is a sharp mode that becomes progressively flatter upon 
increasing field and spreads out over the whole Brillouin zone. In contrast, the 
middle part is dominated by a sharp mode that becomes progressively more 
dispersive upon increasing field and appears to trade intensity with the top 
mode. The lowest energy part is dominated by a continuum spread, with intensity 
moving from bottom to top upon increasing field. All features and trends in the 
INS data are quantitatively captured by exact diagonalization (ED) calculations 
using the full Hamiltonian in (\ref{E:Hamiltonian}) with the expectation values 
$\langle S^y \rangle$ and $\langle S^z \rangle$ in the mean-field term ${\cal 
H}_3$ calculated self-consistently; these calculations are shown in the second 
column of Fig.~\ref{F:expcomparison}.

The breakup of the spectrum in field is clearly illustrated in 
Fig.~\ref{F:expcomparison}I at 1~T. The lowest energy part of the spectrum 
centred around $1.4~$meV shows a set of excitations extending over a broad 
energy range with clear sharp modes visible both at the bottom and the top of 
this range. Those excitations are clearly separated from another set of states 
centred around $2.25$~meV with a clear sharp mode near $2.5$~meV. At higher 
energies still, there is the vestige of the zero-field kinetic bound state near 
$l=1$, now clearly separated from the rest of the spectrum. All the observed 
features, both dispersions and wavevector-dependence of intensities, are well 
captured by the ED calculations in Fig.~\ref{F:expcomparison}J.

The 2.5~T data (Fig.~2Q) shows even more contrasting behaviour between the 
different parts of the spectrum. The sharp mode at the top of the low energy 
continuum now extends all the way from the Brillouin zone center to 
$l\approx0.6$ and has gained in intensity compared to the continuum below it. In 
the middle energy region the sharp mode has become strongly dispersive, with the 
middle continuum losing nearly all its scattering intensity, and the top sharp 
mode has become almost entirely flat and spread out over almost all the 
Brillouin zone. Again, all these features are well reproduced in 
Fig.~\ref{F:expcomparison}R.
The spectra at 0.5~T (Figs.~\ref{F:expcomparison}E and F) and 1.5~T 
(Figs.~\ref{F:expcomparison}M and N) interpolate between 0, 1 and 2.5~T and show 
the gradual evolution of the spectrum.

The ED calculations quantitatively capture every feature and trend described 
above. We stress that the parameters used in this calculation were \emph{not} 
fit to the finite transverse field data presented here, but are fixed to the 
values proposed in \cite{Woodland2023}. This excellent agreement between data 
and calculation gives further support to the Hamiltonian proposed in 
\cite{Woodland2023} and motivates our search for a physical picture of the 
excitations. In the following sections, we will introduce a picture of solitons 
on the zigzag chains and show that the breaking up of the spectrum and very 
different evolution of the different parts in field can be captured 
quantitatively and understood phenomenologically in terms of solitons hopping 
and bound state formation. 

\section{Two-soliton model}\label{S:twosoliton}

In (\ref{E:Hmin}) to (\ref{E:H3}), all $\lambda$ terms are $\ll1$ such that the 
dominant term is the ferromagnetic Ising term. This means that it is sufficient 
for our purposes to consider two-soliton excitations, and neglect mixing with 
four-or-more-soliton excitations, since those occur at much higher energy. More 
systematically, we may write the Hamiltonian as
\begin{equation*}
    \mathcal{H}=\mathcal{H}_{\rm Ising}+\mathcal{V},
\end{equation*}
with $\mathcal{H}_{\rm Ising}$ denoting the Ising Hamiltonian, and $\mathcal{V}$ 
containing all other terms in $\mathcal{H}$. We now treat $\mathcal{V}$ as a 
perturbation of order $\delta$, and define a Schrieffer-Wolff transformation,
\begin{equation*}
 \mathcal{H}^\prime=e^\mathcal{S}\mathcal{H}e^{-\mathcal{S}},
\end{equation*}
with $\mathcal{S}^\dagger=-\mathcal{S}$. We require that
\begin{equation}\label{eq:dwconserve}
 [\mathcal{H}^\prime,\mathcal{H}_{\rm Ising}]=0,   
\end{equation}
i.e., that $\mathcal{H}^\prime$ conserves the number of solitons (domain walls). 
By expanding $\mathcal{S}=\mathcal{S}_1\delta+\mathcal{S}_2\delta^2+...$, and 
enforcing \eqref{eq:dwconserve} up to order $\delta^n$ with $n\in\mathbb{Z}^+$, 
we obtain a systematic perturbative series for the effective Hamiltonian within 
subspaces with a fixed number of solitons. Up to lowest order in $\delta$, we 
arrive at
\begin{equation}\label{eq:Heff}
    \mathcal{H}^\prime\approx\mathcal{H}_{\rm Ising}+\mathcal{V}_{\rm 
conserv}=\sum_{i\geq 0}\mathcal{P}_i\mathcal{H}\mathcal{P}_i,
\end{equation}
with $\mathcal{V}_{\rm conserv}$ denoting the soliton number conserving terms in 
$\mathcal{V}$, and $\mathcal{P}_i$ standing for the projector to the sector with 
exactly $i$ solitons.  We have verified that the second order terms, 
$\sim\delta^2$, are negligible compared to this leading contribution.

Relying on these insights, we now start by considering the effect of the 
effective Hamiltonian ~\eqref{eq:Heff}, first on a single soliton, and then 
within the two-soliton sector. We focus on the minimal Hamiltonian 
$\mathcal{H}_1+\mathcal{H}_3$, yielding a good qualitative understanding of the 
spectrum, and leave the discussion of $\mathcal{H}_2$ to the Appendices.  We 
further assume that the most important contribution from the mean field 
interchain coupling $\mathcal{H}_3$ is a $z$ magnetic field,
\begin{equation*}
    \mathcal{H}_3\approx -h_z\sum_j S_j^z,\quad{\rm with}\; h_z=2J\lambda_{\rm 
MF}\langle S^z\rangle\approx J\lambda_{\rm MF},
\end{equation*}
assuming $\langle S^z\rangle=1/2$.
For convenience, we define the \emph{unconventional} raising and lowering 
operators,
\begin{equation}\label{E:Spm}
S_{j}^\pm = S_{j}^y \mp i S_j^x, 
\end{equation}
which obey the usual commutation relations. Note that this is equivalent to 
performing the calculations in a spin 
basis rotated by $\pi/2$ around the $z$ axis, obtained via the canonical 
transformation $S^x_j\rightarrow -S^y_j$ and $S^y_j\rightarrow S^x_j$ in 
$\mathcal{H}$.

\subsection{Action of the Hamiltonian on a single 
soliton}\label{S:singlesoliton}

In this section we consider the spectrum of deconfined solitons under the 
Hamiltonian $\mathcal{H}_1$ projected to the single soliton sector. The 
confining mean field $\mathcal{H}_3$ will be introduced in 
Sec.~\ref{S:twosolitonsolution}, where we discuss the spectrum in the 
two-soliton sector.

Let us define the ``left'' soliton state, a single domain wall  at the link 
$(j-1,j)$, separating up spins on the left and down spins on the right,
\begin{align}
  \label{eq:2}
  |j\rangle_L & =   \left| 
\cdots\uparrow\uparrow\uparrow_{j-1}\downarrow_j\downarrow\downarrow\cdots\right
\rangle.
\end{align}
Here, the arrows indicate the eigenstates of $S_j^z$ with eigenvalues $\pm 1/2$. 
 We now consider the action of the Hamiltonian $\mathcal{H}_1$ on this state, 
term by term. 
\begin{itemize}
\item {\em Ising exchange:} The single soliton state $|j\rangle_L$ is
  an eigenstate,  with excitation energy
  $\epsilon_0=J/2$ above the ground state.
\item {\em XY exchange $\lambda_S$:} This term flips two adjacent
  spins in opposite directions. Acting on $|j\rangle_L$, it creates two 
additional domain walls by flipping the spins at sites $j-1$ and $j$, and can 
therefore be dropped.
\item {\em transverse field $h_y$:} This term flips a single spin,
\begin{equation*}
    V_{y}=\dfrac{h_y}{2}\sum_j\left(S^+_j+S^-_j\right),
\end{equation*}
where we have used the unconventional raising and lowering operators 
(\ref{E:Spm}).
  To conserve the number of solitons when acting on $|j\rangle_L$, the flipped 
spin must be either at $j$ or $j-1$,
  \begin{align*}
    \mathcal{P}_1 V_{y}|j\rangle_L = \frac{h_y}{2} \left( |j+1\rangle_L 
+|j-1\rangle_L \right).
  \end{align*}
  Therefore, the transverse field gives rise to a nearest neighbour hopping term 
for the domain wall.
\item {\em staggered off-diagonal exchange $\lambda_{yz}$}: Similarly to the 
transverse
  field $h_y$, this term results in a single spin flip,
  \begin{align*}
    V_{yz} & = \dfrac{J\lambda_{yz}}{2}\sum_j  (-1)^j \left( S_{j}^+ + 
S_{j}^-\right) \left( S_{j+1}^z - S_{j-1}^z\right).
  \end{align*}
  Here, the operator $S_{j}^+ + S_{j}^-$ can flip the spin at site $j$ if and 
only if the spins on sites $j+1$ and $j-1$ are opposite, due to the factor 
$S_{j+1}^z-S_{j-1}^z$. Therefore, $V_{yz}$ yields spin flip processes confined 
to domain walls.  We find
   \begin{align*}
    \mathcal{P}_1 V_{yz}|j\rangle_L & = \frac{J\lambda_{yz}}{2} (-1)^{j+1} 
\left(|j+1\rangle_L - |j-1\rangle_L\right),
  \end{align*}
  a hopping term similar to the effect of the field $h_y$, but with a different 
sign structure across links. 
\end{itemize}

To obtain the spectrum of this hopping Hamiltonian, we write a Schr\"odinger 
equation for the soliton. We define the state
\begin{align*}
  |\psi_L\rangle = \sum_{j'} \psi_L(j') |j'\rangle_L,
\end{align*}
where $\psi_L(j)={}_L\langle j|\psi\rangle$ is the wavefunction of
the soliton. Using the expressions derived for $\mathcal{P}_1\mathcal{H}_1 
|j\rangle_L$ above, the Schr\"odinger equation,
  \begin{align*}
    {}_L\langle j|\mathcal{H}_1|\psi_L\rangle = \omega\, \psi_L(j),
  \end{align*}
  can be rewritten as
\begin{align*}
    &\frac{1}{2}\sum_{\Delta=\pm 1}\left(h_y + (-1)^j J\lambda_{yz}\Delta\right) 
\psi_L(j-\Delta) = \left(\omega - \frac{J}{2}\right)\psi_L(j).
  \end{align*}
  This equation describes a staggered hopping of solitons, with hopping 
amplitudes
  \begin{align}\label{eq:hpm}
    h_{\pm}=\dfrac{1}{2}\left(h_y\pm J\lambda_{yz}\right)  
  \end{align}
  alternating on even/odd bonds. 

Since $\mathcal{H}$ is invariant under translations by two lattice sites, we can 
use the following Bloch ansatz,
  \begin{equation*}
    \psi_L(2p+\sigma) = \psi_{L\sigma} e^{ikpc},\quad{\rm with}\; \sigma=0,1.
  \end{equation*}
where $k=2\pi l/c$ is the soliton momentum. In the following, we will 
interchangeably use the symbols $k$ and $l$ when referring to momentum along the 
chain direction, with the only difference that $k$ is in absolute units whereas 
$l$ is in reciprocal lattice units. In the above equation, the coefficients 
$\psi_{L\sigma}$ differentiate between the even and odd sublattices,  reflecting 
the two-site unit cell of the Hamiltonian. From now on, we will reserve the 
index $p$ for labelling the unit cells, whereas $j$ will be used as a label of 
lattice sites. With this convention, the Schr\"odinger equation reduces to
  \begin{align*}
    \left(h_+ e^{-i kc} + h_-  \right)\psi_{L1} & = 
\left(\omega-\frac{J}{2}\right)\psi_{L0}, \\
    \left(h_+ e^{i kc} + h_-  \right)\psi_{L0} & = 
\left(\omega-\frac{J}{2}\right)\psi_{L1},
  \end{align*}
  yielding a pair of bands, $\omega_\pm$, with bonding / anti-bonding orbitals
  \begin{align}
    \label{eq:1dw}
    \left(\frac{J}{2}-\omega_\pm\right)^2 & = \left(h_+ e^{-i kc} + h_-  \right) 
\left( h_+ e^{ikc} +h_-\right) \nonumber \\
    &= h_+^2+h_-^2 +2 h_+ h_- \cos kc.
  \end{align}
  Importantly, the dispersion vanishes if $h_+=0$ or $h_-=0$.  In these limits, 
the hopping amplitude vanishes either on odd or on even bonds, and the domain 
wall can only move between  two sites, giving rise to flat localized bands, as 
previously noted in \cite{Morris2021}. This localized limit will serve as a 
convenient starting point for perturbative considerations in 
Sec.~\ref{S:localized}, allowing us to obtain a simple qualitative picture for 
the evolution of the INS spectrum with magnetic field $h_y$.

  Above we  considered a single ``left'' soliton, describing a domain wall with 
up spins on the left and down spins on the right. Another type of domain wall 
excitation is a ``right'' soliton, separating a domain of down spins on the left 
from up spins on the right,
\begin{equation*}
  |j\rangle_R = \left| 
\cdots\downarrow\downarrow\downarrow_{j-1}\uparrow_j\uparrow\uparrow\cdots\right
\rangle.
\end{equation*}
The arguments described above can be repeated for right solitons, with the only 
difference being  that the hopping induced by $\lambda_{yz}$ is of opposite sign 
compared to the case of left solitons. This interchanges the hopping amplitudes 
$h_+$ and $h_-$ in the Schr\"odinger equations, but leaves the dispersion 
\eqref{eq:1dw} unaltered. Therefore, both solitons become localized at the same 
critical magnetic field $h_y$.

\subsection{Solution of the Hamiltonian in the two-soliton 
subspace}\label{S:twosolitonsolution}

We now turn to the spectrum of the effective Hamiltonian within the two-soliton 
subspace. In Sec.~\ref{S:singlesoliton} we obtained two distinct soliton 
dispersions, $\omega_{\pm}$, describing bonding/anti-bonding orbitals. 
Therefore, we expect three continua in the two-soliton subspace, arising from 
the pairings $(\omega_+,\omega_+)$, $(\omega_+, \omega_-)$ and 
$(\omega_-,\omega_-)$. However, the solitons interact, due both to hard-core 
repulsion and to nearest neighbour soliton-soliton interactions encoded in 
$\mathcal{H}_1$.
Moreover, the full Hamiltonian also includes a confining $z$ magnetic field, 
$\mathcal{H}_3$, yielding an attractive interaction between the two solitons in 
a pair. Below, we take into account all of these effects by considering 
$\mathcal{H}_1+\mathcal{H}_3$ projected to the relevant subspace with 
$\mathcal{P}_2$ and show that a full understanding of the spectrum cannot 
neglect these interactions.

Assuming $h_z>0$, the relevant low energy excitations correspond to a single 
domain of down spins inserted into a background of up spins. Therefore, it is 
convenient to use the following basis,  with a ``left" soliton on the left and a 
``right" soliton on the right,
\begin{equation*}
  |j_L, j_R\rangle =  \left| 
\cdots\uparrow\uparrow_{j_L-1}\downarrow_{j_L}\cdots\downarrow_{j_R-1}
\uparrow_{j_R}\uparrow\cdots\right\rangle,
\end{equation*}
with $j_L<j_R$. Similarly to the procedure followed in 
Sec.~\ref{S:singlesoliton}, we can derive a Schr\"odinger equation within the 
two-soliton subspace by considering the effect of $\mathcal{P}_2\mathcal{H}_1$ 
and $\mathcal{H}_3$ on the basis states.

The transverse field $h_y$ and staggered off-diagonal exchange $\lambda_{yz}$ 
again give rise to hopping terms for the left and right solitons, with a 
correction term arising for nearest neighbor solitons $j_R=j_L+1$ due to hard 
core repulsion,
 \begin{align*}
    \mathcal{P}_2 V_{y}|j_L, j_R\rangle = \frac{h_y}{2} 
\sideset{}{'}\sum_{\Delta=\pm 1}\left( |j_L+\Delta, j_R\rangle 
+|j_L,j_R+\Delta\rangle \right),
  \end{align*}
  and
  \begin{align*}
    &\mathcal{P}_2 V_{yz}|j_L, j_R\rangle =\\
    &\frac{J\lambda_{yz}}{2} \sideset{}{'}\sum_{\Delta=\pm 
1}\Delta\left((-1)^{j_L}|j_L-\Delta, j_R\rangle- 
(-1)^{j_R}|j_L,j_R-\Delta\rangle \right).
  \end{align*}
  Here $\sideset{}{'}\sum$ denotes a restricted summation constrained to valid 
basis states, by dropping the unphysical terms $|j_L+1,j_R\rangle$ and 
$|j_L,j_R-1\rangle$ for neighboring solitons $j_R=j_L+1$.

Besides these familiar terms, two new types of contributions arise compared to 
the single soliton case. The XY exchange term,
\begin{equation*}
V_S=\frac{-J\lambda_S}{2}\sum_{j}\left( S_j^+S_{j+1}^- + S_j^-S_{j+1}^+\right),
\end{equation*}
gives rise to a nearest neighbor interaction term between solitons,
 \begin{align*}
    \mathcal{P}_2 V_S |j_L, j_R\rangle =-\frac{J\lambda_S}{2} 
\delta_{j_R-j_L,1}\sum_{\Delta=\pm 1}|j_L-\Delta, j_R-\Delta\rangle.
  \end{align*}
This term shifts the center of mass coordinate of the soliton pair by one 
lattice site, without changing the relative coordinate $j_R-j_L$. Finally, the 
magnetic field $h_z$ leads to an attractive potential between the left and right 
soliton,
\begin{equation*}
    \mathcal{H}_3|j_L, j_R\rangle =h_z(j_R-j_L)|j_L, j_R\rangle.
\end{equation*}

Based on these relations, we construct the two-particle Schr\"odinger equation 
for the wavefunction $\Psi(j_L,j_R)$ defined through
\begin{equation*}
    |\Psi\rangle = \sum_{j_L<j_R} \Psi(j_L,j_R) |j_L,j_R\rangle.
\end{equation*}
  Relying on translational invariance for the center of mass coordinate, it is 
convenient to write
\begin{equation}\label{eq:Phi}
     \Psi(2p_L+\sigma_L,2p_R+\sigma_R)  =  e^{i k c(p_L+p_R)/2}                  
  \Phi_{\sigma_L\sigma_R}^{(k)}(p_R-p_L),
\end{equation} 
with $p_{L/R}$ labeling the two site unit cells, $\sigma_{L/R}=0,1$ 
distinguishing the even and odd sublattice and ${c(p_L+p_R)/2}$ being the 
position of the center of mass of the soliton pair. For a fixed center of mass 
momentum $k$, we obtain coupled equations for 
$\Phi_{\sigma_L\sigma_R}^{(k)}(n)$, defined on the half line $n\geq 0$ with 
boundary conditions 
$\Phi_{00}^{(k)}(0)=\Phi_{11}^{(k)}(0)=\Phi_{10}^{(k)}(0)=0$. We present more 
details on the numerical solution of these equations in 
Appendix~\ref{A:twosoliton}. 

The two-soliton Schr\"odinger equation derived above yields a spectrum 
consisting of three continua and three bound states across a wide range of 
parameters, see Appendix \ref{A:twosolitonspectrum}. As mentioned above, the 
origin of the three continua can be understood as due to the three different 
ways of combining the two bands of solitons into two-soliton continua. The 
origin of the bound states, which we term $\varepsilon$ bound states to avoid 
confusion, will be explored in the following section, Sec.~\ref{S:localized}. 

Before turning to the detailed study of the $\varepsilon$ bound states, we 
conclude this section by deriving a formula for the INS spectrum within the 
two-soliton model, showing that the dominant contribution stems from the three 
bound states. To this end, we approximate the ground state $|\text{GS}\rangle$ 
appearing in the  dynamical structure factor $S^{xx}$, Eq.~\eqref{E:Sxx}, as the 
ground state of the Ising Hamiltonian, $|\text{GS}\rangle\approx 
|...\uparrow\uparrow\uparrow...\rangle$. Acting with the spin operator $S^x(k)$ 
creates a soliton pair. Using the unconventional raising and lowering operators 
$S_{j}^\pm = S_{j}^y \mp i S_j^x$, we can write the resulting state as 
\begin{equation*}
    S^x(k)|\text{GS}\rangle\approx -\dfrac{i}{2}\sum_j e^{ik j 
c/2}|j,j+1\rangle,
\end{equation*}
where $k=2\pi l/c$ is the soliton pair center of mass momentum. By substituting 
the eigenstates $|\lambda_f\rangle$ with the solutions of the two-soliton 
Schr\"odinger equation constructed above, we arrive at the overlaps
\begin{equation}\label{eq:SQoverlap}
    |\langle\lambda_f|S^{x}(k)|\text{GS}\rangle|^2\sim\left|\Phi_{01}^{(k)}(0)+
    \Phi_{10}^{(k)}(1)\right|^2.
\end{equation}
Therefore, the $\varepsilon$ bound states, which we will show to have a large 
weight on the configurations with nearest neighbor domain walls, give the 
dominant contribution to the dynamical spin structure factor \eqref{E:Sxx}. 

The INS intensity as calculated above is shown in the third column of 
Fig.~\ref{F:expcomparison}. The calculation uses the full Hamiltonian with the 
same parameters as in the ED calculations, except that the spin vector 
expectation value $\langle \mathbf{S}\rangle =(0,0,1/2)$ is assumed fixed, 
rather than using a self-consistent value. This is a good approximation since 
even at 2.5~T, the self-consistent value as calculated by ED is 
$\langle\mathbf{S}\rangle=(0,-0.150,0.473)$. The agreement between the observed 
spectrum and the model is still quantitative --- all features and trends are 
captured --- but not quite as strong as for the exact diagonalization 
calculation. For instance, there is a small overall energy shift, most visible 
by comparing Figs.~\ref{F:expcomparison}R and S; in the latter, energies are 
shifted to lower values. However all key features are well reproduced at all 
measured fields.

To gain more insight into the structure and magnetic field dependence of this 
INS signal, we examine the $\varepsilon$ bound states in the next section, by 
relying on a perturbative argument around the localized limit $h_-=0$.
  
\section{The localized limit}\label{S:localized}
As derived in Sec.~\ref{S:singlesoliton}, the staggered off-diagonal exchange 
$\lambda_{yz}$ and the transverse field $h_y$ in the leading order Hamiltonian 
$\mathcal{H}_1$ lead  to hopping terms of opposite sign for the solitons. A 
particularly interesting situation arises when these terms are matched, such 
that $h_-=0$, resulting in localized single solitons. This localized limit 
serves as a convenient starting point for perturbative considerations, shedding 
light on the structure and magnetic field dependence of the $\varepsilon$ bound 
states, as well as the three two-soliton continua. First, in 
Sec.~\ref{subsec:h-0}, we set  $h_-=0$, and study the two-soliton spectrum, in 
particular, the nature of the two-soliton bound states. We then examine the 
effect of a small non-zero delocalizing term $h_-$ in 
Sec.~\ref{subsec:h-finite}. Predictions for the evolution of the INS spectra 
with decreasing transverse field $h_y$, and comparisons of these to the 
experimental data, are discussed in Sec.~\ref{S:expcomparison}. For most of this 
section, we focus on the leading order Hamiltonian $\mathcal{H}_1$, with a brief 
comment about the mean field $\mathcal{H}_3$ at the end of the section.

\subsection{Localized limit $h_-=0$}\label{subsec:h-0}

\begin{figure}
 {\centering
\includegraphics[width=0.3\textwidth]{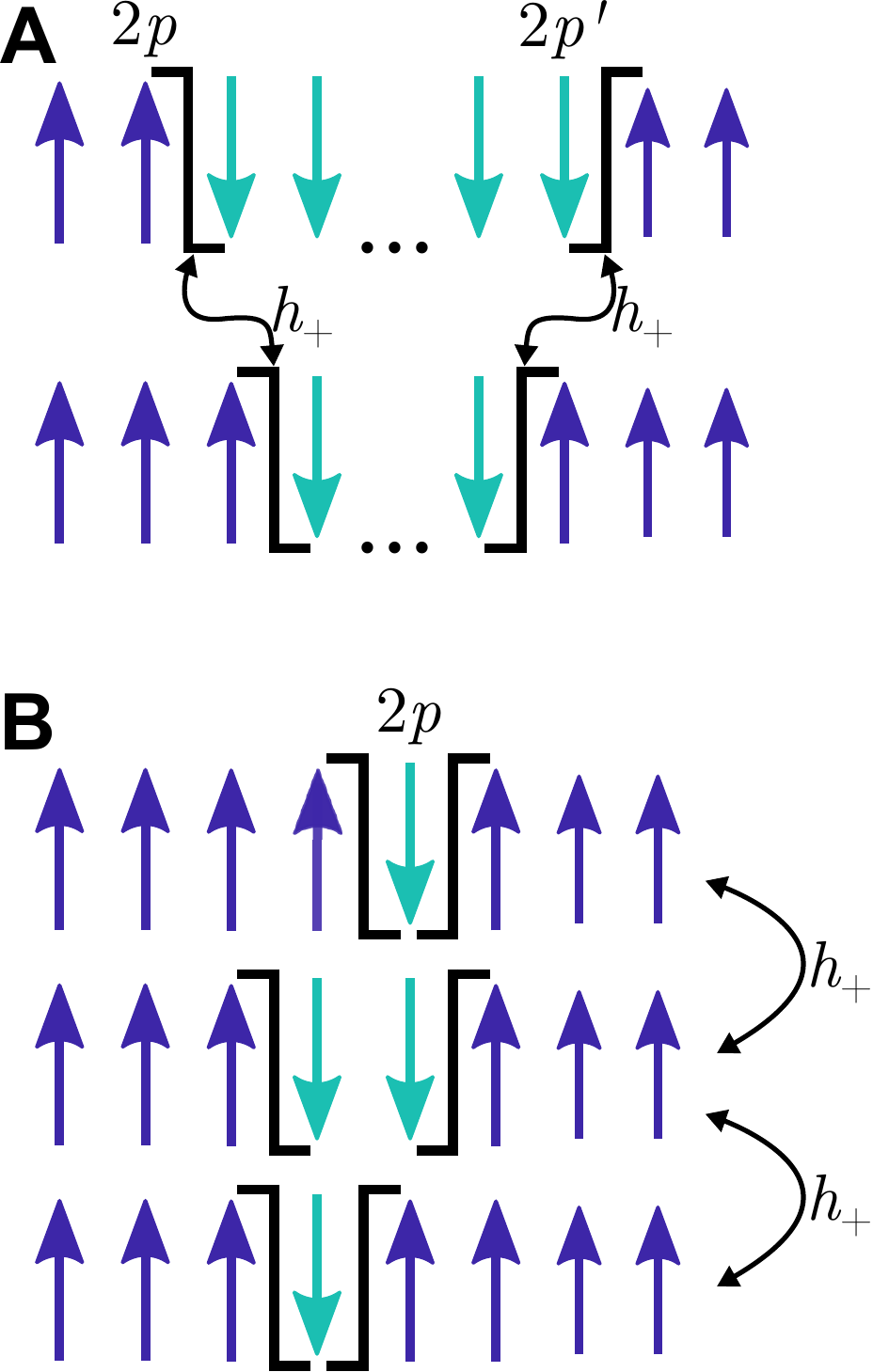}}
 \caption{Soliton hopping in the localized limit. A) In the case of well 
separated solitons, they can hop independently, with each soliton hopping 
between two sites with matrix element $h_+$. B) When solitons are in adjacent 
unit cells, hopping is constrained by hard-core repulsion between solitons.}
\label{F:domainwallhopping}
\end{figure}

For simplicity, we start by setting $\lambda_S=0$ as well, and only keep the 
staggered off-diagonal exchange $\lambda_{yz}$ and the transverse field $h_y$. 
We will discuss the effect of the nearest neighbor exchange $\lambda_S$ later. 
Under these simplifications, a left soliton can hop between the two sites of a 
unit cell $p$, $2p\leftrightarrow 2p+1$, with rate $h_-$, whereas it hops 
between neighboring unit cells $p-1$ and $p$, through sites $2p-1\leftrightarrow 
2p$, with rate $h_+$, see Fig.~\ref{F:domainwallhopping}A. For $h_{-}=0$, we 
obtain the following eigenstates with energies $\omega_\pm$, 
\begin{equation*}
    |p\rangle_L^\pm\equiv\dfrac{1}{\sqrt{2}}\left(|2p-1\rangle_L\pm|2p\rangle_L
    \right),\quad\omega_{\pm}=\frac{J}{2}\pm 
h_+,
\end{equation*}
symmetric and antisymmetric under the inversion exchanging the even and odd 
sublattices, respectively.  For a right soliton, the role of $h_-$ and $h_+$ is 
interchanged, leading to symmetric / antisymmetric eigenstates localized within 
a single unit cell $p$,
\begin{equation*}
    |p\rangle_R^\pm\equiv\dfrac{1}{\sqrt{2}}\left(|2p\rangle_R\pm|2p+1\rangle_R
    \right),\text{ 
with }\omega_{\pm}=\frac{J}{2} \pm h_+.
\end{equation*}

Relying on these observations, we can construct the localized two-soliton 
eigenstates by considering a left soliton confined to sites $2p-1$ and $2p$, and 
a right soliton on $2p^\prime$ and $2p^\prime+1$. For $p^\prime>p$, the solitons 
do not interact, and we obtain the eigenstates
\begin{align*}
    &|p\rangle_L^-\otimes|p^\prime\rangle_R^-, \text{ with energy 
}\omega_{--}=J-2h_+,\\[0.3em]
    &\dfrac{1}{\sqrt{2}}\left[|p\rangle_L^+\otimes|p^\prime\rangle_R^-\pm|p
    \rangle_L^-\otimes|p^\prime\rangle_R^+\right], 
\text{ with }\omega_{+-}=J,\\[0.3em]
    &|p\rangle_L^+\otimes|p^\prime\rangle_R^+, \text{ with }\omega_{++}=J+2h_+.
\end{align*}
The eigenvalue $\omega_{+-}=J$ is doubly degenerate, and the eigenstates were 
chosen to be symmetric / antisymmetric under inversion. We now construct 
delocalized eigenstates with a well defined center of mass momentum $k$ as 
follows
\begin{align}\label{eq:continua}
    &|n,k\rangle^{\pm}=\dfrac{1}{\sqrt{N}}\sum_{\text{cells},p}e^{ik 
c(2p+n)/2}|p\rangle_L^\pm\otimes|p+n\rangle_R^\pm,\nonumber\\
    &|n,k\rangle^0_{\pm}=\dfrac{1}{\sqrt{2N}}\sum_{\text{cells},p}e^{ik 
c(2p+n)/2}\left(|p\rangle_L^+\otimes|p+n\rangle_R^-\right.\nonumber\\
    &\qquad\qquad\qquad\qquad\qquad\quad\left.\pm|p\rangle_L^-\otimes|p+n
    \rangle_R^+
\right),
\end{align}
with $N$ denoting the number of unit cells, and $n\geq 1$. These eigenstates 
correspond to the three two-soliton continua arising from the different pairing 
of bonding / anti-bonding orbitals. In the localized limit considered here, we 
obtain highly degenerate flat bands at energies $J \pm 2 h_+$ and $J$, reflected 
by the free index $n$ standing for the relative coordinate between the left and 
right solitons.

Placing the left soliton to sites $2p-1$ and $2p$, and the right soliton to 
$2p^\prime$ and $2p^\prime+1$ with $p=p^\prime$ gives rise to interaction 
through hard core repulsion, see Fig.~\ref{F:domainwallhopping}B. In this case, 
the eigenstates can be obtained by diagonalizing a $3\times 3$ matrix acting on 
the three allowed configurations, yielding
\begin{align*}
    &|\varepsilon_\pm,p\rangle=\dfrac{1}{2}\left(|2p-1,2p\rangle+|2p,2p+1\rangle
\right.\\
    &\hspace{3cm}\left.\pm\sqrt{2}|2p-1,2p+1\rangle\right),\\
    &|\varepsilon_0,p\rangle=\dfrac{1}{\sqrt{2}}\left(|2p-1,2p\rangle-|2p,2p
+1\rangle\right),
\end{align*}
with energies $\varepsilon_\pm=J\pm\sqrt{2}h_+$ and $\varepsilon_0=J$. The 
corresponding momentum eigenstates  form non-degenerate flat bands 
(Fig.~\ref{F:dispersions_intensity}A) given by
\begin{align}\label{eq:bound}
    &|\varepsilon_\alpha,k\rangle=\dfrac{1}{\sqrt{N}}\sum_{\text{cells},p}e^{ik 
pc}|\varepsilon_\alpha,p\rangle, \text{ for } \alpha=\pm,0.
\end{align}
These states are the origin of the $\varepsilon$ bound states found through the 
numerical solution of the two-soliton Schr\"odinger equation in 
Sec.~\ref{S:twosoliton}.

We now consider the effect of a weak symmetric exchange $\lambda_S$, while 
keeping $h_-=0$. As discussed in Sec.~\ref{S:twosoliton}, this term only affects 
nearest neighbor solitons by shifting the center of mass coordinate. Therefore, 
the three continua, \eqref{eq:continua}, with solitons residing in different 
unit cells are not affected. In contrast, the symmetric exchange acts 
non-trivially on the bound states ~\eqref{eq:bound}, inducing an energy shift 
calculated perturbatively as
\begin{align*}
    \varepsilon_\alpha\longrightarrow\varepsilon_\alpha(k)\approx\varepsilon_
    \alpha+
\langle 
\varepsilon_\alpha,k |\,V_S\,|\varepsilon_\alpha,k\rangle,
\end{align*}
yielding dispersive bands
\begin{align}\label{eq:e_pert_ls}
    &\varepsilon_\pm(k)=J\pm\sqrt{2}h_+-\dfrac{J\lambda_S}{4}\left(1+\cos{kc}
\right),\nonumber\\
    &\varepsilon_0(k)=J+\dfrac{J\lambda_S}{2}\left(1+\cos{kc}\right).
\end{align}

The full spectrum of the localized limit $h_-=0$, with weak symmetric exchange 
$\lambda_S$ is illustrated in Fig.~\ref{F:dispersions_intensity}B, showing the 
three highly degenerate flat continua, and the three dispersive $\varepsilon$ 
bound states, which become delocalized by $\lambda_S$. Note that in 
\ch{CoNb2O6}, $\lambda_S$ is large enough that the dispersion causes the 
$\varepsilon_0$ and $\varepsilon_+$ bands to cross, leading to band inversion. 
This effect is discussed quantitatively in Appendix \ref{S:bandinversion} and 
illustrated in Fig.~\ref{F:dispersions_intensity}C. The band inversion further 
suppresses the dispersion of the top mode, as well as mixing the character of 
the two bands. 

\subsection{Effects of weak delocalizing hopping $h_-$}\label{subsec:h-finite}

\begin{figure*}
    \includegraphics[width=\textwidth]{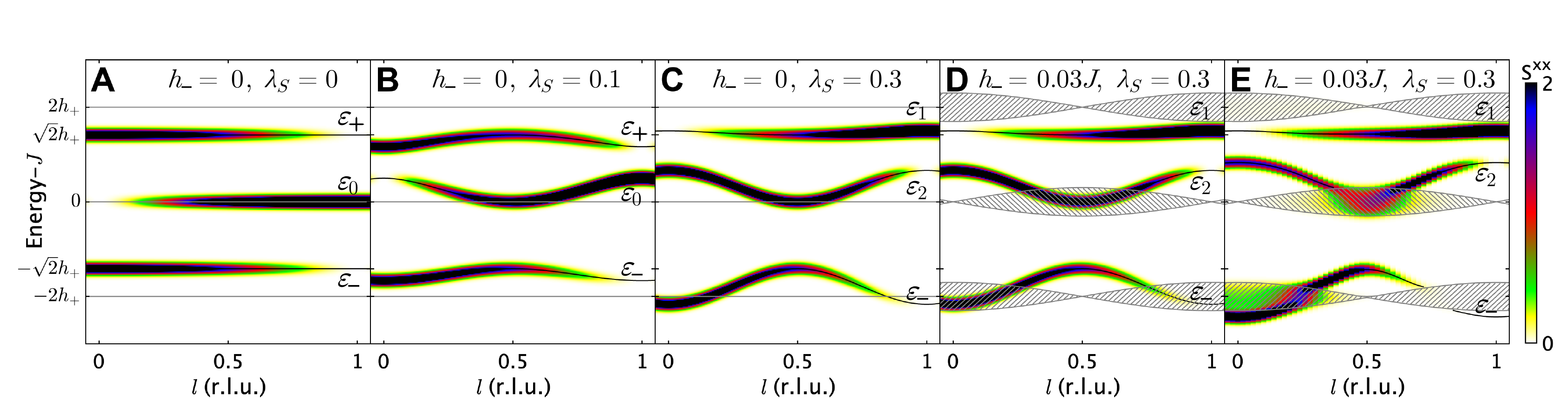}
    \caption{Dispersions and intensities as a function of wavevector and energy 
at different stages of perturbation away from the purely localized limit. The 
$x$-axis is momentum in reciprocal lattice units, $l=kc/(2\pi)$, vertical axis 
is energy relative to $J$. In each panel, color indicates $S^{xx}$, as defined 
in (\ref{E:Sxx}) and further normalized by the total number of sites, on a 
linear scale indicated by the colorbar; black lines represent $\varepsilon$ 
bound states while gray hatched patches represent continua. The calculated 
intensities have been convolved with a Gaussian of FWHM $0.033J$ and are shown 
in absolute units of $1/J$. In all panels, $h_+=0.2J$. In A-D) the calculation 
uses the perturbative regime of Sec.~\ref{S:localized}. A) Bound states and 
continua are both dispersionless for $h_-=0$ and $\lambda_S=0$. B) For finite 
$\lambda_S$, bound states become dispersive but continua remain flat and highly 
degenerate.  C) Band inversion occurs between the top two bound states 
$\varepsilon_+ $ and $\varepsilon_0$ for $J\lambda_S > 2\sqrt{2}h_{+}/3$ (see 
Appendix~\ref{S:bandinversion}), resulting in modes labelled 
$\varepsilon_{1,2}$, such that $\varepsilon_1$ inherits the structure of the 
state $\varepsilon_0$ close to $l=0$ and $l=1$ while retaining the character of 
$\varepsilon_+$ elsewhere, with $\varepsilon_2$ showing the opposite behavior. 
This is reflected by the change in intensity distribution from B to C near 
$l=0,1$. D) Continua become dispersive for finite $h_-$. E) Same as D, but 
calculated using the full two-soliton model in Sec.~\ref{S:twosoliton}, which 
predicts hybridization between the lowest two bound states, $\varepsilon_2$ and 
$\varepsilon_-$, and the continua they overlap.}
    \label{F:dispersions_intensity}
\end{figure*}

As shown above, the localized limit $h_-=0$ provides remarkable insight into the 
structure of the two-soliton spectrum obtained near to this limit. We can gain a 
more detailed understanding of the evolution of the spectrum with decreasing 
transverse field by considering  the effect of a weak delocalizing hopping term 
$h_-$ perturbatively. The first important effect of $h_-\neq 0$ is lifting the 
high degeneracy of the three continua and broadening these bands. Secondly, a 
finite $h_-$ can mix the $\varepsilon$ bound states constructed in the previous 
section with the continua in $k$ regions where they overlap in energy.

We first explore the first effect, by focusing on the continua, and applying 
degenerate perturbation theory within each band separately. We note that for 
weak $h_-$ and $\lambda_S$, the bottom and top bands remain well separated from 
the $\varepsilon$ bound states. Assuming also $\lambda_S\gg h_-$, the middle 
continuum overlaps with the middle bound state only in the vicinity of 
$k=\pi/c$, see Fig.~\ref{F:dispersions_intensity}D. Therefore, our treatment of 
neglecting the hybridization between the continua and the $\varepsilon$ bound 
states is justified in the limit of weak couplings $h_+\gg\lambda_S\gg h_-$, 
apart from the case of the middle band in the vicinity of $k=\pi/c$. While the 
experimental parameters lie outside of this well controlled region, we will 
demonstrate below that a first order perturbative expansion grants valuable 
insight into the evolution of the INS spectra for the whole range of applied 
transverse fields.

Denoting the hopping term by $V_{h_-}$, we find that the matrix elements between 
the single soliton eigenstates of the localized limit are
\begin{align}\label{eq:Vhmin}
    _\alpha^\pm\langle 
p^\prime|V_{h_-}|p\rangle_\alpha^\pm=\pm\dfrac{h_-}{2}\left(\delta_{p^\prime,p
+1}+\delta_{p^\prime,p-1}\right),\;\; 
\alpha=L,R.
\end{align}
Up to first order in perturbation theory, the energy shifts of the three 
continua can be evaluated by considering the matrix elements of $V_{h_-}$ 
between the two-soliton eigenstates of the localized limit, \eqref{eq:continua}, 
within each band separately. 
Relying on the relations~\eqref{eq:Vhmin}, we obtain for the top and bottom 
bands
\begin{align*}
  ^{\pm}\langle n^\prime,k^\prime|V_{h_-}&|n,k\rangle^{\pm}=\\
  &\pm \delta_{k,k^\prime}\, 
h_-\cos\left(\frac{kc}{2}\right)\left(\delta_{n,n^\prime+1}+\delta_{n,n^
\prime-1}\right).
\end{align*}
For a fixed center of mass momentum $k$, this equation corresponds to an 
effective  nearest neighbor hopping Hamiltonian for the relative coordinate $n$, 
with hopping amplitude $\pm h_-\cos(kc/2)$, subject to the hard core constraint 
$n>0$. Therefore, for the bottom and top bands we find the spectrum,
\begin{equation*}
   \omega_{\sigma\sigma} = J+2\sigma \left(h_+ +  h_- 
\cos\left(\frac{kc}{2}\right)\cos\left(\frac{qc}{2}\right)\right), \;\; 
\sigma=\pm1,
\end{equation*}
with approximate unnormalized eigenstates
\begin{equation*}
    |q,k\rangle^\pm\sim\sum_{\substack{\text{cell separation},\\ 
n>0}}\sin\left(\frac{qnc}{2}\right)\,|n,k\rangle^\pm.
\end{equation*}
Here, $k$ is the total momentum of the soliton pair and $q$ is the relative 
momentum of the two solitons, and the factor $\sin(qnc/2)$ reflects hard core 
repulsion $n>0$. Thus, the weak hopping $h_-$ broadens the bands, most strongly 
around $k=0$, but the high degeneracy still persists at $k=\pi/c$, see 
Fig.~\ref{F:dispersions_intensity}D. 

Turning to the middle continuum, we have to calculate four types of matrix 
elements between the two-soliton states $|n,k\rangle_\pm^0$. We obtain
\begin{align*}
  & _-^0\langle n^\prime,k^\prime|V_{h_-}|n,k\rangle_+^0=\, _+^0\langle 
n^\prime,k^\prime|V_{h_-}|n,k\rangle_-^0 \\
  & \hspace{2.cm} =\delta_{k,k^\prime}\,i\, 
h_-\sin\left(\frac{kc}{2}\right)\left(\delta_{n^\prime,n+1}-\delta_{n^
\prime,n-1}\right), 
\\[0.4em]
  & _-^0\langle n^\prime,k^\prime|V_{h_-}|n,k\rangle_-^0=\, _+^0\langle 
n^\prime,k^\prime|V_{h_-}|n,k\rangle_+^0 =0.
\end{align*}
 We can now calculate the broadening of this band by diagonalizing this  matrix 
within a given total momentum sector $k$. We find that the eigenstates in the 
middle band remain at least twofold degenerate everywhere, yielding the spectrum
\begin{equation*}
    \omega_{+-}=J-2 h_- 
\sin\left(\frac{kc}{2}\right)\sin\left(\frac{qc}{2}\right),
\end{equation*}
with approximate unnormalized eigenstates
\begin{align*}
    &|q,k\rangle^0_1\sim\!\sum_{\substack{\text{cell separation},\\ 
n>0}}\!\!\left( e^{iqcn/2}-e^{-i(qc/2+\pi)n} \right)|n,k\rangle^0_{\xi(n)},\\
    &|q,k\rangle^0_2\sim\!\sum_{\substack{\text{cell separation},\\ 
n>0}}\!\!\left( e^{iqcn/2}-e^{-i(qc/2+\pi)n} \right)|n,k\rangle^0_{\xi(n+1)},
\end{align*}
 with $\xi(n) = \pm$ for $n$ even/odd, and $q$ again standing for the relative 
momentum. Thus, the middle band remains highly degenerate around $k=0$, but it 
is broadened away from this point, most strongly around $k=\pi/c$.

We note that these dispersions of the continua can be understood based on the 
single-soliton spectrum as given in \eqref{eq:1dw}.
In the limit of small $h_-$, \eqref{eq:1dw} becomes
\begin{equation}\label{eq:1dwlocalized}
\omega_{\pm}=\frac{J}{2}\pm(h_+ + h_-\cos kc).
\end{equation}
The three continua constructed above correspond to the three  different types of 
soliton pairs. Denoting the individual solitons' momenta by $k_1$ and $k_2$, we 
obtain the  energies
\begin{align*}
\omega_{++}=&J+2h_+ + h_-(\cos k_1 c+\cos k_2 c)\nonumber \\
=& J+2h_+ + 2h_-\cos\left(\frac{kc}{2}\right)\cos\left(\frac{qc}{2}\right), 
\nonumber\\
\omega_{--}=&J-2h_+ - h_-(\cos k_1 c+\cos k_2 c)\nonumber \\
=& J-2h_+ - 2h_-\cos\left(\frac{kc}{2}\right)\cos\left(\frac{qc}{2}\right), 
\nonumber\\
\omega_{+-}=&J + h_-(\cos k_1 c-\cos k_2 c)\nonumber \\
=& J - 2h_-\sin\left(\frac{kc}{2}\right)\sin\left(\frac{qc}{2}\right),
\end{align*}
with total wavevector $k=k_1+k_2$, and relative momentum $q=k_1-k_2$, in 
accordance with the expressions derived above.

 The second important effect of the hopping $h_-$ is mixing the $\varepsilon$ 
bound states with the continua where they overlap in energy. In these regions, 
the originally bound soliton pair can become delocalized, broadening out the INS 
signal, as shown in Fig.~\ref{F:dispersions_intensity}E. 

\subsection{Comparison with INS data}\label{S:expcomparison}

We conclude this section by describing the INS intensity predicted by these 
perturbative arguments. In the localized limit $h_-=0$, the continua 
\eqref{eq:continua} have no overlap with nearest neighbor soliton pairs 
$|j,j+1\rangle$ so do not contribute to the overlap \eqref{eq:SQoverlap} and to 
the resulting INS spectrum. The $\varepsilon$ bound states \eqref{eq:bound}, on 
the other hand, have the property that \begin{align*}
    &\langle\varepsilon_\pm,k|S^x(k)|\text{GS}\rangle = 
-i\sqrt{N}\frac{e^{ikc/2}+1}{4},\\
    &\langle\varepsilon_0,k|S^x(k)|\text{GS}\rangle= 
-i\sqrt{N}\frac{e^{ikc/2}-1}{2\sqrt{2}},
\end{align*}
following from comparing \eqref{eq:bound} to \eqref{eq:Phi}. Therefore, the 
different bound states contribute differently to the INS spectrum 
$S^{xx}(\mathbf{Q},\omega)$. This structure is inherited by the bound states 
away from the limit $h_-=0$.

In the absence of band inversion ($J\lambda_S<2\sqrt{2}h_+/3$), the three bound 
states can be labeled by $\varepsilon_{\pm,0}$, such that the bottom and top 
modes $\varepsilon_\pm$ yield strong signals in the vicinity of $l=0$ and are 
suppressed near $l=1$ [note that $l=kc/(2\pi)$], whereas the middle band 
$\varepsilon_0$ behaves in the opposite way, see 
Fig.~\ref{F:dispersions_intensity}B. If $\lambda_S$ is large enough for band 
inversion to occur, the top and middle bands acquire new labels 
$\varepsilon_{1,2}$, with $\varepsilon_1$ retaining the structure of 
$\varepsilon_+$ around $l=0.5$ but inheriting the character of $\varepsilon_0$ 
around $l=0$ and $l=1$, and the reverse holding for $\varepsilon_2$. This 
results in a transfer of intensity between the top two bound states in the 
vicinity of $l=0$ and $l=1$. That is, the top mode is strong at $l=1$ and weak 
at $l=0$, and this is opposite to the middle band 
(Fig.~\ref{F:dispersions_intensity}C).

For small but finite $h_-$, the lowest $\varepsilon$ bound state mode 
$\varepsilon_-$, which is strong around $l=0$ in the limit $h_-=0$, is pushed 
into the bottom continuum due to the strong broadening of the continuum with 
$h_-$. The mixing between these states leads to a signal smeared out across a 
larger range of energies. Similarly, the middle bound state mode mixes with the 
middle continuum around $l=0.5$, smearing and eventually almost completely 
washing out the signal, see Fig.~\ref{F:dispersions_intensity}E.
However, for $\lambda_S$ large enough to lead to band inversion, the top mode 
does not hybridize with the continuum around $l=1$, even for large $h_-$. This 
is because the upper continuum consists of states that are even under exchange 
of the even and odd sublattices, whereas the bound state $\varepsilon_0$ is odd, 
shedding light on the remarkably sharp INS spectrum of the top state around 
$l=1$, even far from the localized limit $h_-=0$. That is, the top mode 
$\varepsilon_1$ is sharp at all fields in Fig.~\ref{F:expcomparison} in the data 
(first column) and in the calculation (third column), even though there are 
regions where it overlaps with states originating from the top continuum, as 
illustrated in Fig.~\ref{F:expcomparison}, last column. 

This right-most column of Fig.~\ref{F:expcomparison} shows the INS intensity as 
calculated in this section, using the same parameters as for the middle two 
columns but with no longitudinal mean field ($\mathcal{H}_3=0$). Comparing 
Fig.~\ref{F:expcomparison}T and P with Q and M respectively, it is seen that the 
above description is in remarkable qualitative agreement with the experimental 
results in large fields 2.5~T and 1.5~T. The agreement is especially good at 
2.5~T (compare Figs.~\ref{F:expcomparison}S and T), which is close to the 
localized limit. This agreement is strongly supportive of the model presented in 
this section, especially given that the model is relatively simple and 
completely analytically tractable. At lower fields, the agreement is expected to 
be less good as the solitons are now more delocalized and so a perturbative 
treatment around the localized limit is expected to be less quantitatively 
accurate.
Nevertheless, several key trends are still reproduced: in particular, the top 
mode $\varepsilon_1$ is captured at all fields, the middle mode $\varepsilon_2$ 
is captured down to 1~T, and the lowest mode $\varepsilon_{-}$ is captured down 
to 1.5~T.

The calculations in the right-most column of Fig.~\ref{F:expcomparison} include 
the effect of band inversion on the top two bound states, as well as the terms 
in $\mathcal{H}_2$, as explained in Appendix \ref{S:bandinversion}. We note that 
the full Hamiltonian also contains a $z$ magnetic field, $\mathcal{H}_3$, 
changing the nature of the continua. This term introduces a linear confinement, 
splitting the continua into confinement bound states, which cannot be captured 
in this calculation. Despite these important effects, the tightly bound 
$\varepsilon$ bound states remain relatively unaffected by the confinement, and 
the qualitative predictions for the INS spectra presented in this section still 
hold, see Appendix~\ref{A:twosolitonspectrum}. We note that, at first order, a 
finite longitudinal mean field ($\mathcal{H}_3$) would be expected to increase 
the energies of all the $\varepsilon$-bound states, which would bring the 
calculated dispersions in the right-most column of Fig. \ref{F:expcomparison}  
into closer agreement with the experimental data (left-most column).

\section{Conclusions}\label{S:conclusion}
We investigated the spectrum of the Ising chain material \ch{CoNb2O6} as a 
function of low to intermediate transverse field in the ordered phase using 
inelastic neutron scattering experiments. We compared the measured spectrum to 
predictions based on a recently refined Hamiltonian containing all relevant 
sub-leading terms beyond the dominant Ising exchange and found strong 
quantitative agreement. We then sought a physical picture of the excitations. We 
found that by restricting the Hilbert space to the two-soliton subspace at first 
order in perturbation theory, very good agreement between the calculation and 
experiment was still achieved. The resulting spectrum in general has three 
continua and three bound states, of which only the bound states contribute 
significant weight to the inelastic neutron scattering intensity. In order to 
understand the character of the bound states, we considered the localized limit, 
in which the soliton hopping term on alternate bonds is zero. This occurs when 
the applied field matches the strength of the off-diagonal exchange. We found 
that the bound states in this limit are of two solitons in adjacent unit cells, 
stabilized by hardcore repulsion leading to a change in delocalization energy. 
The bound states survive well away from the localized limit, suggesting that 
this picture has a broader domain of validity than might initially be expected.
Using this physical picture, we have been able to gain both qualitative and 
quantitative understanding of the low energy spectrum of \ch{CoNb2O6} in the low 
transverse field ordered phase.

\begin{acknowledgments}
L.W. acknowledges support from a doctoral studentship funded by Lincoln College 
and the University of Oxford. I.L. acknowledges support from the Gordon and 
Betty Moore Foundation through Grant GBMF8690 to UCSB and from the National 
Science Foundation under Grant No. NSF PHY-1748958. D.P. acknowledges support 
from the Engineering and Physical Sciences Research Council grant number 
GR/M47249/01. L.B. was supported by the NSF CMMT program under Grant No. 
DMR-2116515, and by the Simons Collaboration on Ultra-Quantum Matter, which is a 
grant from the Simons Foundation (651440). R.C. acknowledges support from the 
European Research Council under the European Union’s Horizon 2020 research and 
innovation programme Grant Agreement Number 788814 (EQFT). The neutron 
scattering measurements at the ISIS Facility were supported by a beamtime 
allocation from the Science and Technology Facilities Council. Access to the 
data will be made available from Ref.~\cite{database}.
\end{acknowledgments}

\appendix

\section{Solution of the two-soliton Schr\"odinger equation}\label{A:twosoliton}

In this Appendix we present more details on the derivation and numerical 
solution of the two-soliton Schr\"odinger equation constructed in 
Sec.~\ref{S:twosoliton}. Using the action of $\mathcal{P}_2\mathcal{H}_1$ and 
$\mathcal{H}_3$ on the basis states, the Schr\"odinger equation,
\begin{equation*}
    \langle j_L, 
j_R|\mathcal{H}_1\mathcal{P}_2+\mathcal{H}_3|\Psi\rangle=\omega\Psi(j_L,j_R),
\end{equation*}
yields
\begin{widetext}
\begin{align*}
    \dfrac{1}{2} \sideset{}{'}\sum_{\Delta=\pm 
1}\left[\left(h_y+\Delta(-1)^{j_L}J\lambda_{yz}\right) \Psi(j_L-\Delta,j_R) 
+\left(h_y-\Delta(-1)^{j_R}J\lambda_{yz}\right)\Psi(j_L,j_R-\Delta)\right] &\\
    -\frac{J\lambda_S}{2} \delta_{j_R-j_L,1}\sum_{\Delta=\pm 
1}\Psi(j_L-\Delta,j_R-\Delta)+h_z(j_R-j_L)\Psi(j_L,j_R)&=(\omega-2\epsilon_0)
\Psi(j_L,j_R).
  \end{align*}
  Here $\epsilon_0=J/2$ is the energy cost of a single domain wall, and 
$\sideset{}{'}\sum$ stands for a constrained summation restricted to the 
physical domain of $\Psi(j_L^\prime,j_R^\prime)$, $j_L^\prime<j_R^\prime$. 
Introducing the center of mass momentum $k$, and rewriting this equation in 
terms of $\Phi_{\sigma_L\sigma_R}^{(k)}(n)$ according to \eqref{eq:Phi}, with 
$n$ labeling the distance between two-site unit cells, leads to
  \begin{align*}
  h_+ e^{-ikc/2}\Phi_{10}^{(k)}(n+&1)+h_- \Phi_{10}^{(k)}(n) + h_- e^{-i kc/2} 
\Phi _{01}^{(k)}(n-1) + h_+ \Phi_{01}^{(k)}(n)+2 n h_z \Phi_{0,0}^{(k)}(n)= 
(\omega-2\epsilon_0) \Phi_{00}^{(k)}(n), &&n\geq 1,\\[0.3em]
 & h_- \Phi_{00}^{(k)}(n) + h_+ e^{i kc/2} \Phi_{00}^{(k)}(n-1) + h_- e^{-i 
kc/2} \Phi_{11}^{(k)}(n-1)\\
  &\qquad+ h_+ \Phi_{11}^{(k)}(n) + h_z(2n-1) 
\Phi_{10}^{(k)}(n)-J\lambda_S\,\delta_{n,1}\cos 
\left(\frac{kc}{2}\right)\Phi_{01}^{(k)}(n-1) = 
(\omega-2\epsilon_0)\Phi_{10}^{(k)}(n),&& n\geq 1,\\[0.3em]
 & h_+ e^{-i kc/2} \Phi_{11}^{(k)}(n+1)+h_- \Phi_{11}^{(k)}(n) + h_+ 
\Phi_{00}^{(k)}(n)+ h_- e^{i kc/2} \Phi_{00}^{(k)}(n+1)\\
 & \qquad\qquad+h_z(2n+1) \Phi_{01}^{(k)}(n)- J\lambda_S\, \delta_{n,0} \cos 
\left(\frac{kc}{2}\right) \Phi_{10}^{(k)}(n+1) =  (\omega-2\epsilon_0) 
\Phi_{01}^{(k)}(n), && n\geq 0,\\[0.3em]
 &\hspace{-2.3cm} h_- \Phi_{01}^{(k)}(n)+ h_+ e^{i kc/2} \Phi_{01}^{(k)}(n-1) + 
h_+ \Phi_{10}^{(k)}(n) + h_- e^{i kc/2} \Phi_{10}^{(k)}(n+1)+2n 
h_z\Phi_{11}^{(k)}(n) =  (\omega-2\epsilon_0) \Phi_{11}^{(k)}(n), &&n\geq 1,
  \end{align*}
\end{widetext}
with boundary conditions 
$\Phi_{00}^{(k)}(0)=\Phi_{11}^{(k)}(0)=\Phi_{10}^{(k)}(0)=0$. This set of 
equations can be solved numerically by truncating them at a large maximal 
distance between the solitons, $n_{\rm max}$ or by using periodic boundary 
conditions on a finite ring. By defining the vector

\begin{equation*}
  \bm{\Phi}^{(k)} =  
    \begin{pmatrix}
    \Phi_{01}^{(k)}(0) \\[0.5em]
    \Phi_{00}^{(k)}(1) \\[0.5em]
    \Phi_{11}^{(k)}(1) \\[0.5em]
    \Phi_{10}^{(k)}(1)\\[0.5em]
    \Phi_{01}^{(k)}(1) \\[0.5em]
    \Phi_{00}^{(k)}(2) \\[0.5em]
    \Phi_{11}^{(k)}(2) \\
    \vdots
  \end{pmatrix},
\end{equation*}
we obtain a matrix equation, allowing us to determine the low energy spectrum.

\section{Effect of subleading terms $\mathcal{H}_2$ in the two-soliton 
picture}\label{app:subleading}

In this Appendix we briefly discuss the correction terms to the two-soliton 
Schr\"odinger equation derived in Appendix~\ref{A:twosoliton} arising from the 
subleading couplings in $\mathcal{H}_2$. We examine the action of 
$\mathcal{H}_2$ term by term. 
Following the convention of Sec.~\ref{S:twosoliton}, we rewrite $\mathcal{H}_2$ 
in a  rotated basis, $S^x_j\rightarrow -S^y_j$ and $S^y_j\rightarrow S^x_j$.

 We first consider the anti-symmetric nearest neighbor coupling,
 \begin{equation}\label{eq:VA}
V_A=\frac{J\lambda_A}{2}\sum_{\text{sites},j}\left( S_j^+S_{j+1}^+ + 
S_j^-S_{j+1}^-\right),
\end{equation}
raising or lowering two neighboring spins. Projected to the single soliton 
subspace, this term moves the soliton by two sites,
\begin{equation*}
    \mathcal{P}_1 
V_A|j\rangle_L=\dfrac{J\lambda_A}{2}\left(|j-2\rangle_L+|j+2\rangle_L\right).
\end{equation*}
In the two-soliton region, we obtain a next to nearest neighbor hopping term for 
the left and right solitons, whenever permitted by the hard core constraint 
$j_R>j_L$. Applying the representation~\eqref{eq:Phi}, we get the following 
contribution to the left hand side of the Schr\"odinger equation for 
$(\omega-2\epsilon_0)\Phi_{\sigma_L\sigma_R}^{(k)}(n)$,
\begin{align*}
J\lambda_A&\cos\left(\frac{kc}{2}\right)\left[\Phi_{\sigma_L\sigma_R}^{(k)}(n+1)
\right.\\
&\left.+(1-\delta_{n,0}-\delta_{n,1}(1-\delta_{\sigma_L,0}\delta_{\sigma_R,1}))
\,\Phi_{\sigma_L\sigma_R}^{(k)}(n-1)\right].
\end{align*}

The perturbation
\begin{equation}\label{eq:VAFxy}
V_{\text{AF}}^{xy}=\frac{J\lambda_{\text{AF}}^{xy}}{2}\sum_{\text{sites},j}
\left(S_j^+S_{j+2}^- 
+ S_j^-S_{j+2}^+\right)
\end{equation}
flips a pair of next nearest neighbor spins in opposite directions. Acting on a 
single soliton, $V_{\rm AF}^{xy}|j\rangle_L$ always leaves the single soliton 
subspace. In the presence of two solitons, $|j_L,j_R\rangle$, however, we get a 
non-vanishing short range contribution for $j_R\leq j_L+2$,
\begin{align*}
    \mathcal{P}_2 V_{\rm AF}^{xy}|j_L,j_R\rangle = \dfrac{J\lambda_{\rm 
AF}^{xy}}{2}&\sum_{\Delta=\pm 
1}\left(\delta_{j_R,j_L+1}\,|j_L+2\Delta,j_R+2\Delta\rangle\right. \\
    &\qquad+\left.\delta_{j_R,j_L+2}\,|j_L+\Delta,j_R+\Delta\rangle\right).
\end{align*}
This term shifts the center of mass coordinate by $\pm 2$ or $\pm 1$ sites for a 
spin down domain of length $j_R-j_L=1$ and $j_R-j_L=2$, respectively. In the 
Schr\"odinger equation, it leads to the following four extra contributions on 
the left hand side,
\begin{align*}
    J\lambda_{\rm AF}^{xy}\cos(kc)\Phi_{01}^{(k)}(0)&\longleftarrow 
(\omega-2\epsilon_0)\Phi_{01}^{(k)}(0), \\[0.3em]
    J\lambda_{\rm AF}^{xy}\cos(kc)\Phi_{10}^{(k)}(1)&\longleftarrow 
(\omega-2\epsilon_0)\Phi_{10}^{(k)}(1), \\[0.3em]
   J\lambda_{\rm AF}^{xy}\frac{1+e^{-ikc}}{2}\Phi_{11}^{(k)}(1)&\longleftarrow 
(\omega-2\epsilon_0)\Phi_{00}^{(k)}(1), \\[0.3em]
   J\lambda_{\rm AF}^{xy}\frac{1+e^{ikc}}{2}\Phi_{00}^{(k)}(1)&\longleftarrow 
(\omega-2\epsilon_0)\Phi_{11}^{(k)}(1).
\end{align*}

Finally, the effect of the perturbation 
\begin{equation}\label{eq:VAF}
V_{\text{AF}}=J\lambda_{\text{AF}}\sum_{\text{sites},j}S_j^zS_{j+2}^z
\end{equation}
is to lower the energy of all states relative to the fully aligned (ground) 
state. This term is diagonal in the two-soliton basis $|j_L,j_R\rangle$, 
yielding an energy shift depending on the size of the spin down domain. If there 
is only a single spin flip, $j_R-j_L=1$, only two antiferromagnetic bonds are 
satisfied, whereas if there are two or more spin flips, $j_R-j_L\geq 2$, four 
antiferromagnetic bonds are satisfied, i.e.,
\begin{align*}
V_{\text{AF}}|j,j+1\rangle&=-J\lambda_{\text{AF}}|j,j+1\rangle \nonumber\\
V_{\text{AF}}|j,j+2\rangle&=-2J\lambda_{\text{AF}}|j,j+2\rangle,
\end{align*}
where only the energy difference between the excited state and the ground state 
has been kept.  
These considerations lead to the energy shift
\begin{align*}
    &\omega-2\epsilon_0 \longrightarrow \omega-2\epsilon_0 -J\lambda_{\rm 
AF},\quad {\rm if}\;\; 2n+\sigma_R-\sigma_L=1, \\
    &\omega-2\epsilon_0 \longrightarrow \omega-2\epsilon_0 -2J\lambda_{\rm 
AF},\quad {\rm if}\;\; 2n+\sigma_R-\sigma_L>1,
\end{align*}
on the left hand side of the Schr\"odinger equation for 
$\Phi_{\sigma_L\sigma_R}^{(k)}(n)$.

\section{Bound states in the two-soliton spectrum}\label{A:twosolitonspectrum}

\begin{figure}
\includegraphics[width=0.5\textwidth]{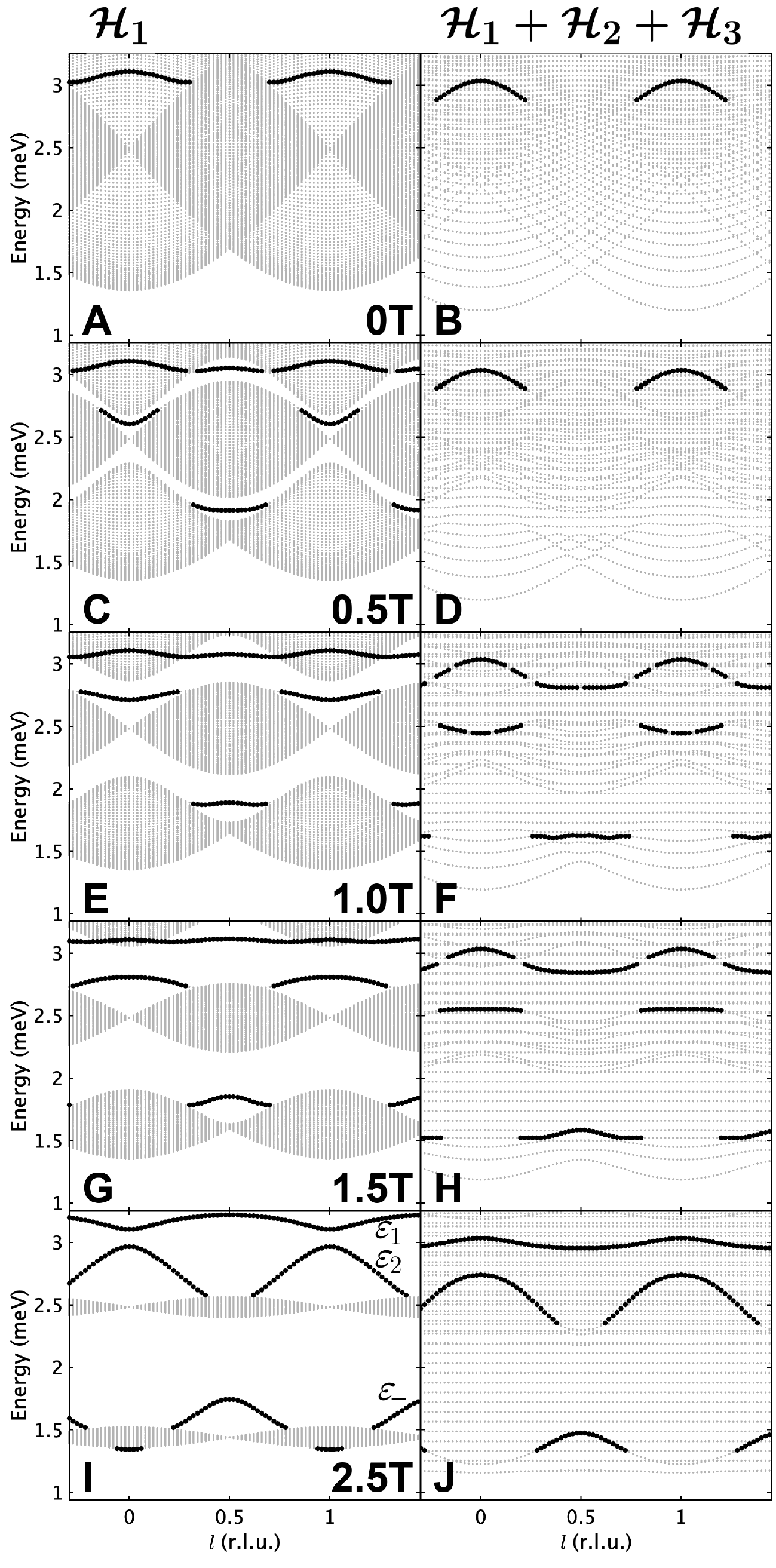}
\caption{Eigenstates of the two-soliton Hamiltonian as a function of wavevector 
and energy at different fields, increasing from top to bottom. In each panel, 
states with the character of the $\varepsilon$ bound states have been 
highlighted in black. These states have been identified by being well separated 
from other states in energy, or, where they overlap with other states, by having 
high INS intensity and discarding regions with strong hybridization.  Left 
column: Solution of $\mathcal{H}_1$. Three continua and three bound states can 
be seen at every non-zero field. $\varepsilon_1$, $\varepsilon_2$  and 
$\varepsilon_-$ are defined in Appendix~\ref{S:bandinversion}. Right column: 
solution of $\mathcal{H}=\mathcal{H}_1+\mathcal{H}_2+\mathcal{H}_3$. This 
comparison illustrates the key effect of $\mathcal{H}_3$, i.e., the presence of 
a longitudinal mean field: all continua are split into confinement bound states, 
but the $\varepsilon$ bound states are left essentially intact. This column is 
to be compared with the third column of Fig.~\ref{F:expcomparison}, which shows 
the intensities under the same conditions.}\label{F:twosoliton}
\end{figure}

In this Appendix, the two-soliton spectrum in various regimes is briefly 
discussed. The left column of Fig.~\ref{F:twosoliton} shows the two-soliton 
spectrum as a function of field for the minimal Hamiltonian $\mathcal{H}_1$. In 
this case, the spectrum consists of three continua and three bound states, whose 
origins are discussed in Sec.~\ref{S:localized}, at every non-zero field. The 
right hand column of Fig.~\ref{F:twosoliton} shows the same calculation for the 
full Hamiltonian $\mathcal{H}_1+\mathcal{H}_2+\mathcal{H}_3$, where 
$\mathcal{H}_3$ is the confining mean field. In this case, the mean field splits 
the continua into a series of confinement bound states, but the $\varepsilon$ 
bound states are left mostly intact, because they are tightly bound.

\section{Bound state inversion and matrix elements of $\mathcal{H}_2$ in the 
localized limit}\label{S:bandinversion}
In this Appendix, we consider the spectrum near the localized limit, $h_-=0$, 
obtained when the transverse field satisfies $h_y=J\lambda_{yz}$. When 
$3J\lambda_S/(2\sqrt{2}) > h_+$, as is the case for the experimentally relevant 
parameters, the top two $\varepsilon$ bound state modes cross each other, so we 
must use degenerate perturbation theory within the subspace of these top two 
modes to calculate the resulting spectrum.

The unperturbed bound states and their energies are given in \eqref{eq:bound}. 
In the following, we consider the effects of various other terms in the 
Hamiltonian in (\ref{E:Hmin}) to (\ref{E:H3}), starting with the second term in 
$\mathcal{H}_1$.
We consider the matrix elements between the two highest energy modes 
($|\varepsilon_+, k\rangle$ and $|\varepsilon_0, k\rangle$) for the perturbation
\begin{equation*}
V_S=\frac{-J\lambda_S}{2}\sum_{\text{sites},j} S_j^+S_{j+1}^- + S_j^-S_{j+1}^+.
\end{equation*}
This perturbation allows single spin flips to hop by one site.
The diagonal matrix elements are
\begin{align*}
\langle \varepsilon_+, k |V_S|\varepsilon_+, k'\rangle =& 
\frac{-J\lambda_S}{4}(1+\cos kc)\,\delta_{k,k'} \nonumber\\
\langle \varepsilon_0, k |V_S|\varepsilon_0, k'\rangle =& 
\frac{+J\lambda_S}{2}(1+\cos kc)\,\delta_{k,k'} ,
\end{align*}
and are consistent with the expressions given in Sec. \ref{S:localized}.
Within the degenerate subspace, off-diagonal matrix elements are
\begin{equation*}
\langle 
\varepsilon_0,k|V_S|\varepsilon_+,k'\rangle=\frac{iJ\lambda_S}{2\sqrt{2}}
\sin(kc)\,\delta_{k,k'}
\end{equation*}
and Hermitian conjugate. 
Eigenvalues and eigenvectors are then obtained by direct diagonalization, with 
the dynamical correlations $S^{xx}$ obtained from the eigenvectors as described 
in Sec. \ref{S:expcomparison}. We term the resulting modes $\varepsilon_1$ and 
$\varepsilon_2$.

We also derive the matrix elements of the next nearest neighbour terms in 
$\mathcal{H}_2$. Consistent with the considerations above, we only keep the 
diagonal matrix elements between bound states of the same type, as well as the 
off-diagonal matrix elements between the middle and top modes $\varepsilon_0$ 
and $\varepsilon_+$, which show a strong mixing at the experimental parameters.
The effect of the perturbation $V_{\text{AF}}$, \eqref{eq:VAF},
is to lower the energy of all states relative to the fully aligned (ground) 
state. However, as discussed in Appendix ~\ref{app:subleading}, this energy 
shift is different for states with a single spin flip compared to states with at 
least two spin flips, corresponding to two and four satisfied antiferromagnetic 
bonds, respectively.
This leads to matrix elements
\begin{align*}
\langle \varepsilon_+,k|V_{\text{AF}}|\varepsilon_+,k'\rangle=\langle 
\varepsilon_-,k|V_{\text{AF}}|\varepsilon_-,k'\rangle=&-\frac{3}{2}J\lambda_{\text{AF}}\delta_{k,k'} 
\nonumber\\
\langle 
\varepsilon_0,k|V_{\text{AF}}|\varepsilon_0,k'\rangle=&-J\lambda_{\text{AF}}\delta_{k,k'} 
\nonumber\\
\langle \varepsilon_0,k|V_{\text{AF}}|\varepsilon_+,k'\rangle=&\;0.
\end{align*}
This perturbation also shifts the energies of all the continua by 
$-2J\lambda_{\text{AF}}$.

The effect of the perturbation $V_{\text{AF}}^{xy}$, \eqref{eq:VAFxy},
is to hop single spin flips by two sites. This leads to matrix elements
\begin{align*}
\langle \varepsilon_+,k|V_{\text{AF}}^{xy}|\varepsilon_+,k'\rangle=\langle 
\varepsilon_-,k|V_{\text{AF}}^{xy}|\varepsilon_-,k'\rangle=&\frac{J
\lambda_{\text{AF}}^{xy}}{2}\cos(kc)\delta_{k,k'} 
\nonumber\\
\langle 
\varepsilon_0,k|V_{\text{AF}}^{xy}|\varepsilon_0,k'\rangle=&J
\lambda_{\text{AF}}^{xy}\cos(kc)\delta_{k,k'}\nonumber\\
\langle \varepsilon_0,k|V_{\text{AF}}^{xy}|\varepsilon_+,k'\rangle=&\;0.
\end{align*}
To first order, this perturbation vanishes when acting on the continua. However, 
it mixes the continua with the bound states.

The interchain mean field term, $\mathcal{H}_3$, is
\begin{equation*}
V_z= -h_z\sum_{\text{sites},j}S_j^z
\end{equation*}
under the approximation that $\langle\mathbf{S}\rangle=(0,0,1/2)$.
The effect of this term on the bound states is determined by the matrix elements 
\begin{align*}
\langle \varepsilon_+,k|V_z|\varepsilon_+,k'\rangle=\langle 
\varepsilon_-,k|V_z|\varepsilon_-,k'\rangle=& \frac{3}{2}h_z\delta_{k,k'} 
\nonumber\\
\langle \varepsilon_0,k|V_z|\varepsilon_0,k'\rangle=&h_z\delta_{k,k'}\nonumber\\
\langle \varepsilon_0,k|V_z|\varepsilon_+,k'\rangle=&\;0.
\end{align*}
The effect of this term on the continua is to confine the soliton pairs into a 
series of bound states; this effect cannot be captured within the current 
picture.

Finally we note that the first term in $\mathcal{H}_2$, $V_A$, \eqref{eq:VA},
vanishes when projected to the subspace of bound states, since this term causes 
solitons to hop by two sites at a time. To understand the effect of this term on 
the continua, we note that $V_A$, a hopping term to a neighboring unit cell, 
shifts the single soliton dispersion relations as  
\begin{equation*}
    \omega_\pm\longrightarrow \omega_\pm + J\lambda_A\cos(kc).
\end{equation*}
In contrast to the effect of the hopping $h_-$, $V_A$ induces the same shift in 
the energies of bonding / antibonding orbitals. As a result, this perturbation 
leads to the same energy change for the three continua,
\begin{align*}
    \omega_{\sigma,\sigma^\prime}&\longrightarrow\omega_{\sigma,\sigma^\prime}+J
\lambda_A\left[\cos(k_1c)+\cos(k_2c)\right]\\
    &=\omega_{\sigma,\sigma^\prime}+J\lambda_A\cos\left(\dfrac{kc}{2}\right)\cos
    \left(\dfrac{qc}{2}\right),
\end{align*}
for $\sigma,\sigma^\prime=\pm$, with total and relative momenta given by 
$k=k_1+k_2$ and $q=k_1-k_2$, respectively. 

Alternatively, these spectra can be obtained by applying first order 
perturbation theory around the localized limit, similarly to the analysis of the 
hopping $h_-$ presented in the main text. To this end, we first evaluate the 
effect of $V_A$ on the states $|n,k\rangle$ constructed in 
Sec.~\ref{subsec:h-finite}. We find
\begin{align*}
&^{\pm}\langle n^{\prime}, k^{\prime}|V_A|n,k\rangle^{\pm}= ^{0}_{\pm}\langle 
n^{\prime}, k^{\prime}|V_A|n,k\rangle^{0}_{\pm}  \\
&\hspace{2cm} 
=J\lambda_A\cos\left(\frac{kc}{2}\right)\left(\delta_{n^{\prime},n+1}+
\delta_{n^{\prime},n-1}\right)\delta_{k,k^{\prime}},
\end{align*}
corresponding to an effective nearest neighbor hopping Hamiltonian for the 
relative coordinate $n$, with hopping amplitude $J\lambda_A\cos(kc/2)$, subject 
to the hard core constraint $n>0$.
For the top and bottom continua, the hopping amplitude due to $V_{h_-}$ is also 
real, so the $|q,k\rangle^{\pm}$ states constructed in 
Sec.~\ref{subsec:h-finite} are also eigenstates of $V_A$ and the change in the 
energies is
\begin{equation*}
\frac{^\pm\langle q,k|V_A|q,k\rangle^\pm}{^\pm\langle q,k|q,k\rangle^\pm} = 
2J\lambda_A\cos\left(\frac{kc}{2}\right)\cos\left(\frac{qc}{2}\right).
\end{equation*}
For the middle continuum, we consider  the effect of ${V_{h_-}+V_A}$ on the 
plane wave 
\begin{equation*}
\sum_n e^{iqcn/2}(|n,k\rangle^0_++|n,k\rangle^0_-).
\end{equation*}
 We find that the perturbation corresponds to an effective nearest neighbour 
hopping Hamiltonian for the relative coordinate $n$, with complex hopping 
amplitude $t=J\lambda_A\cos(kc/2)+i 
h_-\sin(kc/2)=t^{\prime}+it^{\prime\prime}=|t|e^{i\varphi}$.
This yields the dispersion 
\begin{align*}
&\omega^A_{+-} = J+2|t|\cos\left(\frac{qc}{2}+\varphi\right)\\
& 
=J+2t^{\prime}\cos\left(\frac{qc}{2}\right)-2t^{\prime\prime}\sin\left(
\frac{qc}{2}\right)=\\
&J-2h_-\sin\left(\frac{kc}{2}\right)\sin\left(\frac{qc}{2}\right)+2J\lambda_A
\cos\left(\frac{kc}{2}\right)\cos\left(\frac{qc}{2}\right).
\end{align*}
The eigenstates satisfying the hard core repulsion boundary condition at $n=0$ 
can be obtained by noting that the plane wave with relative momentum $q$ is 
degenerate with the plane wave with relative momentum $-q-4\varphi/c$. Mixing 
these plane waves leads to the unnormalized eigenstates satisfying the hard core 
constraint,
\begin{equation*}
\sum_{\substack{\text{cell separation},\\ n>0}} 
\left(e^{iqcn/2}-e^{-i(qc/2+2\varphi)n}\right)\left(|n,k\rangle^0_++
|n,k\rangle^0_-\right).
\end{equation*}
For the plane wave
\begin{equation*}
\sum_n e^{iqcn/2}(|n,k\rangle^0_+-|n,k\rangle^0_-),
\end{equation*}
the effect of $V_{h_-}+V_A$ corresponds to an effective nearest neighbour 
hopping Hamiltonian for the relative coordinate $n$, with complex hopping 
amplitude $t^*$, such that the argument above applies with $\varphi \rightarrow 
-\varphi$. Thus, the effect of the perturbation $V_A$ is to add 
$2J\lambda_A\cos(kc/2)\cos(qc/2)$ to the energies of all continua, as 
anticipated based on the single soliton dispersion relation. We also note that 
the argument above yields the following two degenerate eigenstates in the 
presence of $V_{h_-}$ but without $V_A$, i.e. for $\varphi=\pi/2$,
\begin{equation*}
\sum_{ n>0} \left(e^{icq_\pm n/2}- e^{-i(cq_\pm 
/2+\pi)n}\right)\left(|n,k\rangle^0_+\pm |n,k\rangle^0_-\right),
\end{equation*}
with $q_+-q_-=\pi$. The eigenstates constructed in the main text  are the 
symmetric / antisymmetric combinations of these eigenstates.

For $\lambda_A<0$ and $h_->0$ such as is found experimentally, the perturbation 
$V_A$ leads the top continuum to narrow and the bottom continuum to broaden, and 
the middle continuum to broaden around what would otherwise be the nodes. The 
plots in the right-most column of Fig.~\ref{F:expcomparison} include the effects 
of all terms in $\mathcal{H}_1$ and $\mathcal{H}_2$, but not $\mathcal{H}_3$ 
since it is not possible to include the effects of this last term on the 
continua in this framework.


\begin{thebibliography}{20}%
\makeatletter
\providecommand \@ifxundefined [1]{%
 \@ifx{#1\undefined}
}%
\providecommand \@ifnum [1]{%
 \ifnum #1\expandafter \@firstoftwo
 \else \expandafter \@secondoftwo
 \fi
}%
\providecommand \@ifx [1]{%
 \ifx #1\expandafter \@firstoftwo
 \else \expandafter \@secondoftwo
 \fi
}%
\providecommand \natexlab [1]{#1}%
\providecommand \enquote  [1]{``#1''}%
\providecommand \bibnamefont  [1]{#1}%
\providecommand \bibfnamefont [1]{#1}%
\providecommand \citenamefont [1]{#1}%
\providecommand \href@noop [0]{\@secondoftwo}%
\providecommand \href [0]{\begingroup \@sanitize@url \@href}%
\providecommand \@href[1]{\@@startlink{#1}\@@href}%
\providecommand \@@href[1]{\endgroup#1\@@endlink}%
\providecommand \@sanitize@url [0]{\catcode `\\12\catcode `\$12\catcode
  `\&12\catcode `\#12\catcode `\^12\catcode `\_12\catcode `\%12\relax}%
\providecommand \@@startlink[1]{}%
\providecommand \@@endlink[0]{}%
\providecommand \url  [0]{\begingroup\@sanitize@url \@url }%
\providecommand \@url [1]{\endgroup\@href {#1}{\urlprefix }}%
\providecommand \urlprefix  [0]{URL }%
\providecommand \Eprint [0]{\href }%
\providecommand \doibase [0]{https://doi.org/}%
\providecommand \selectlanguage [0]{\@gobble}%
\providecommand \bibinfo  [0]{\@secondoftwo}%
\providecommand \bibfield  [0]{\@secondoftwo}%
\providecommand \translation [1]{[#1]}%
\providecommand \BibitemOpen [0]{}%
\providecommand \bibitemStop [0]{}%
\providecommand \bibitemNoStop [0]{.\EOS\space}%
\providecommand \EOS [0]{\spacefactor3000\relax}%
\providecommand \BibitemShut  [1]{\csname bibitem#1\endcsname}%
\let\auto@bib@innerbib\@empty
\bibitem [{\citenamefont {Sachdev}(1999)}]{Sachdev1999}%
  \BibitemOpen
  \bibfield  {author} {\bibinfo {author} {\bibfnamefont {S.}~\bibnamefont
  {Sachdev}},\ }\href@noop {} {\emph {\bibinfo {title} {Quantum Phase
  Transitions}}}\ (\bibinfo  {publisher} {Cambridge University Press},\
  \bibinfo {address} {London},\ \bibinfo {year} {1999})\BibitemShut {NoStop}%
\bibitem [{\citenamefont {Pfeuty}(1970)}]{Pfeuty1970}%
  \BibitemOpen
  \bibfield  {author} {\bibinfo {author} {\bibfnamefont {P.}~\bibnamefont
  {Pfeuty}},\ }\bibfield  {title} {\bibinfo {title} {The one-dimensional
  {I}sing model with a transverse field},\ }\href
  {https://doi.org/https://doi.org/10.1016/0003-4916(70)90270-8} {\bibfield
  {journal} {\bibinfo  {journal} {Ann. Phys. (NY)}\ }\textbf {\bibinfo {volume}
  {57}},\ \bibinfo {pages} {79} (\bibinfo {year} {1970})}\BibitemShut {NoStop}%
\bibitem [{\citenamefont {Lieb}\ \emph {et~al.}(1961)\citenamefont {Lieb},
  \citenamefont {Schultz},\ and\ \citenamefont {Mattis}}]{Lieb1961}%
  \BibitemOpen
  \bibfield  {author} {\bibinfo {author} {\bibfnamefont {E.}~\bibnamefont
  {Lieb}}, \bibinfo {author} {\bibfnamefont {T.}~\bibnamefont {Schultz}},\ and\
  \bibinfo {author} {\bibfnamefont {D.}~\bibnamefont {Mattis}},\ }\bibfield
  {title} {\bibinfo {title} {Two soluble models of an antiferromagnetic
  chain},\ }\href
  {https://doi.org/https://doi.org/10.1016/0003-4916(61)90115-4} {\bibfield
  {journal} {\bibinfo  {journal} {Ann. Phys. (NY)}\ }\textbf {\bibinfo {volume}
  {16}},\ \bibinfo {pages} {407} (\bibinfo {year} {1961})}\BibitemShut
  {NoStop}%
\bibitem [{\citenamefont {McCoy}\ and\ \citenamefont {Wu}(1978)}]{McCoy1978}%
  \BibitemOpen
  \bibfield  {author} {\bibinfo {author} {\bibfnamefont {B.~M.}\ \bibnamefont
  {McCoy}}\ and\ \bibinfo {author} {\bibfnamefont {T.~T.}\ \bibnamefont {Wu}},\
  }\bibfield  {title} {\bibinfo {title} {Two-dimensional {I}sing field theory
  in a magnetic field: Breakup of the cut in the two-point function},\ }\href
  {https://doi.org/10.1103/PhysRevD.18.1259} {\bibfield  {journal} {\bibinfo
  {journal} {Phys. Rev. D}\ }\textbf {\bibinfo {volume} {18}},\ \bibinfo
  {pages} {1259} (\bibinfo {year} {1978})}\BibitemShut {NoStop}%
\bibitem [{\citenamefont {Coldea}\ \emph {et~al.}(2010)\citenamefont {Coldea},
  \citenamefont {Tennant}, \citenamefont {Wheeler}, \citenamefont {Wawrzynska},
  \citenamefont {Prabhakaran}, \citenamefont {Telling}, \citenamefont
  {Habicht}, \citenamefont {Smeibidl},\ and\ \citenamefont
  {Kiefer}}]{Coldea2010}%
  \BibitemOpen
  \bibfield  {author} {\bibinfo {author} {\bibfnamefont {R.}~\bibnamefont
  {Coldea}}, \bibinfo {author} {\bibfnamefont {D.~A.}\ \bibnamefont {Tennant}},
  \bibinfo {author} {\bibfnamefont {E.~M.}\ \bibnamefont {Wheeler}}, \bibinfo
  {author} {\bibfnamefont {E.}~\bibnamefont {Wawrzynska}}, \bibinfo {author}
  {\bibfnamefont {D.}~\bibnamefont {Prabhakaran}}, \bibinfo {author}
  {\bibfnamefont {M.}~\bibnamefont {Telling}}, \bibinfo {author} {\bibfnamefont
  {K.}~\bibnamefont {Habicht}}, \bibinfo {author} {\bibfnamefont
  {P.}~\bibnamefont {Smeibidl}},\ and\ \bibinfo {author} {\bibfnamefont
  {K.}~\bibnamefont {Kiefer}},\ }\bibfield  {title} {\bibinfo {title} {Quantum
  criticality in an {I}sing chain: Experimental evidence for emergent {E8}
  symmetry},\ }\href {https://doi.org/10.1126/science.1180085} {\bibfield
  {journal} {\bibinfo  {journal} {Science}\ }\textbf {\bibinfo {volume}
  {327}},\ \bibinfo {pages} {177} (\bibinfo {year} {2010})}\BibitemShut
  {NoStop}%
\bibitem [{\citenamefont {Kinross}\ \emph {et~al.}(2014)\citenamefont
  {Kinross}, \citenamefont {Fu}, \citenamefont {Munsie}, \citenamefont
  {Dabkowska}, \citenamefont {Luke}, \citenamefont {Sachdev},\ and\
  \citenamefont {Imai}}]{Kinross2014xe}%
  \BibitemOpen
  \bibfield  {author} {\bibinfo {author} {\bibfnamefont {A.~W.}\ \bibnamefont
  {Kinross}}, \bibinfo {author} {\bibfnamefont {M.}~\bibnamefont {Fu}},
  \bibinfo {author} {\bibfnamefont {T.~J.}\ \bibnamefont {Munsie}}, \bibinfo
  {author} {\bibfnamefont {H.~A.}\ \bibnamefont {Dabkowska}}, \bibinfo {author}
  {\bibfnamefont {G.~M.}\ \bibnamefont {Luke}}, \bibinfo {author}
  {\bibfnamefont {S.}~\bibnamefont {Sachdev}},\ and\ \bibinfo {author}
  {\bibfnamefont {T.}~\bibnamefont {Imai}},\ }\bibfield  {title} {\bibinfo
  {title} {Evolution of quantum fluctuations near the quantum critical point of
  the transverse field {I}sing chain system {C}o{N}b$_2${O}$_6$},\ }\href
  {https://doi.org/10.1103/PhysRevX.4.031008} {\bibfield  {journal} {\bibinfo
  {journal} {Phys. Rev. X}\ }\textbf {\bibinfo {volume} {4}},\ \bibinfo {pages}
  {031008} (\bibinfo {year} {2014})}\BibitemShut {NoStop}%
\bibitem [{\citenamefont {Morris}\ \emph {et~al.}(2014)\citenamefont {Morris},
  \citenamefont {Vald\'es~Aguilar}, \citenamefont {Ghosh}, \citenamefont
  {Koohpayeh}, \citenamefont {Krizan}, \citenamefont {Cava}, \citenamefont
  {Tchernyshyov}, \citenamefont {McQueen},\ and\ \citenamefont
  {Armitage}}]{Morris2014ux}%
  \BibitemOpen
  \bibfield  {author} {\bibinfo {author} {\bibfnamefont {C.~M.}\ \bibnamefont
  {Morris}}, \bibinfo {author} {\bibfnamefont {R.}~\bibnamefont
  {Vald\'es~Aguilar}}, \bibinfo {author} {\bibfnamefont {A.}~\bibnamefont
  {Ghosh}}, \bibinfo {author} {\bibfnamefont {S.~M.}\ \bibnamefont
  {Koohpayeh}}, \bibinfo {author} {\bibfnamefont {J.}~\bibnamefont {Krizan}},
  \bibinfo {author} {\bibfnamefont {R.~J.}\ \bibnamefont {Cava}}, \bibinfo
  {author} {\bibfnamefont {O.}~\bibnamefont {Tchernyshyov}}, \bibinfo {author}
  {\bibfnamefont {T.~M.}\ \bibnamefont {McQueen}},\ and\ \bibinfo {author}
  {\bibfnamefont {N.~P.}\ \bibnamefont {Armitage}},\ }\bibfield  {title}
  {\bibinfo {title} {Hierarchy of bound states in the one-dimensional
  ferromagnetic {I}sing chain {C}o{N}b$_2${O}$_6$ investigated by
  high-resolution time-domain terahertz spectroscopy},\ }\href
  {https://doi.org/10.1103/PhysRevLett.112.137403} {\bibfield  {journal}
  {\bibinfo  {journal} {Phys. Rev. Lett.}\ }\textbf {\bibinfo {volume} {112}},\
  \bibinfo {pages} {137403} (\bibinfo {year} {2014})}\BibitemShut {NoStop}%
\bibitem [{\citenamefont {Liang}\ \emph {et~al.}(2015)\citenamefont {Liang},
  \citenamefont {Koohpayeh}, \citenamefont {Krizan}, \citenamefont {McQueen},
  \citenamefont {Cava},\ and\ \citenamefont {Ong}}]{Liang2015}%
  \BibitemOpen
  \bibfield  {author} {\bibinfo {author} {\bibfnamefont {T.}~\bibnamefont
  {Liang}}, \bibinfo {author} {\bibfnamefont {S.~M.}\ \bibnamefont
  {Koohpayeh}}, \bibinfo {author} {\bibfnamefont {J.~W.}\ \bibnamefont
  {Krizan}}, \bibinfo {author} {\bibfnamefont {T.~M.}\ \bibnamefont {McQueen}},
  \bibinfo {author} {\bibfnamefont {R.~J.}\ \bibnamefont {Cava}},\ and\
  \bibinfo {author} {\bibfnamefont {N.~P.}\ \bibnamefont {Ong}},\ }\bibfield
  {title} {\bibinfo {title} {Heat capacity peak at the quantum critical point
  of the transverse {I}sing magnet {C}o{N}b$_2${O}$_6$},\ }\href
  {https://doi.org/10.1038/ncomms8611} {\bibfield  {journal} {\bibinfo
  {journal} {Nat. Commun.}\ }\textbf {\bibinfo {volume} {6}},\ \bibinfo {pages}
  {7611} (\bibinfo {year} {2015})}\BibitemShut {NoStop}%
\bibitem [{\citenamefont {Amelin}\ \emph {et~al.}(2020)\citenamefont {Amelin},
  \citenamefont {Engelmayer}, \citenamefont {Viirok}, \citenamefont {Nagel},
  \citenamefont {R\~{o}\~{o}m}, \citenamefont {Lorenz},\ and\ \citenamefont
  {Wang}}]{Amelin2020ov}%
  \BibitemOpen
  \bibfield  {author} {\bibinfo {author} {\bibfnamefont {K.}~\bibnamefont
  {Amelin}}, \bibinfo {author} {\bibfnamefont {J.}~\bibnamefont {Engelmayer}},
  \bibinfo {author} {\bibfnamefont {J.}~\bibnamefont {Viirok}}, \bibinfo
  {author} {\bibfnamefont {U.}~\bibnamefont {Nagel}}, \bibinfo {author}
  {\bibfnamefont {T.}~\bibnamefont {R\~{o}\~{o}m}}, \bibinfo {author}
  {\bibfnamefont {T.}~\bibnamefont {Lorenz}},\ and\ \bibinfo {author}
  {\bibfnamefont {Z.}~\bibnamefont {Wang}},\ }\bibfield  {title} {\bibinfo
  {title} {Experimental observation of quantum many-body excitations of {E8}
  symmetry in the {I}sing chain ferromagnet {C}o{N}b$_2${O}$_6$},\ }\href
  {https://doi.org/10.1103/PhysRevB.102.104431} {\bibfield  {journal} {\bibinfo
   {journal} {Phys. Rev. B}\ }\textbf {\bibinfo {volume} {102}},\ \bibinfo
  {pages} {104431} (\bibinfo {year} {2020})}\BibitemShut {NoStop}%
\bibitem [{\citenamefont {Maartense}\ \emph {et~al.}(1977)\citenamefont
  {Maartense}, \citenamefont {Yaeger},\ and\ \citenamefont
  {Wanklyn}}]{Maartense1977ef}%
  \BibitemOpen
  \bibfield  {author} {\bibinfo {author} {\bibfnamefont {I.}~\bibnamefont
  {Maartense}}, \bibinfo {author} {\bibfnamefont {I.}~\bibnamefont {Yaeger}},\
  and\ \bibinfo {author} {\bibfnamefont {B.~M.}\ \bibnamefont {Wanklyn}},\
  }\bibfield  {title} {\bibinfo {title} {Field-induced magnetic transitions of
  {C}o{N}b$_2${O}$_6$ in ordered state},\ }\href@noop {} {\bibfield  {journal}
  {\bibinfo  {journal} {Solid State Commun.}\ }\textbf {\bibinfo {volume}
  {21}},\ \bibinfo {pages} {93} (\bibinfo {year} {1977})}\BibitemShut {NoStop}%
\bibitem [{\citenamefont {Scharf}\ \emph {et~al.}(1979)\citenamefont {Scharf},
  \citenamefont {Weitzel}, \citenamefont {Yaeger}, \citenamefont {Maartense},\
  and\ \citenamefont {Wanklyn}}]{Scharf1979gi}%
  \BibitemOpen
  \bibfield  {author} {\bibinfo {author} {\bibfnamefont {W.}~\bibnamefont
  {Scharf}}, \bibinfo {author} {\bibfnamefont {H.}~\bibnamefont {Weitzel}},
  \bibinfo {author} {\bibfnamefont {I.}~\bibnamefont {Yaeger}}, \bibinfo
  {author} {\bibfnamefont {I.}~\bibnamefont {Maartense}},\ and\ \bibinfo
  {author} {\bibfnamefont {B.~M.}\ \bibnamefont {Wanklyn}},\ }\bibfield
  {title} {\bibinfo {title} {Magnetic-structures of {C}o{N}b$_2${O}$_6$},\
  }\href {https://doi.org/https://doi.org/10.1016/0304-8853(79)90044-1}
  {\bibfield  {journal} {\bibinfo  {journal} {J. Magn. Magn. Mater.}\ }\textbf
  {\bibinfo {volume} {13}},\ \bibinfo {pages} {121} (\bibinfo {year}
  {1979})}\BibitemShut {NoStop}%
\bibitem [{\citenamefont {Mitsuda}\ \emph {et~al.}(1994)\citenamefont
  {Mitsuda}, \citenamefont {Hosoya}, \citenamefont {Wada}, \citenamefont
  {Yoshizawa}, \citenamefont {Hanawa}, \citenamefont {Ishikawa}, \citenamefont
  {Miyatani}, \citenamefont {Saito},\ and\ \citenamefont {Kohn}}]{Mitsuda1994}%
  \BibitemOpen
  \bibfield  {author} {\bibinfo {author} {\bibfnamefont {S.}~\bibnamefont
  {Mitsuda}}, \bibinfo {author} {\bibfnamefont {K.}~\bibnamefont {Hosoya}},
  \bibinfo {author} {\bibfnamefont {T.}~\bibnamefont {Wada}}, \bibinfo {author}
  {\bibfnamefont {H.}~\bibnamefont {Yoshizawa}}, \bibinfo {author}
  {\bibfnamefont {T.}~\bibnamefont {Hanawa}}, \bibinfo {author} {\bibfnamefont
  {M.}~\bibnamefont {Ishikawa}}, \bibinfo {author} {\bibfnamefont
  {K.}~\bibnamefont {Miyatani}}, \bibinfo {author} {\bibfnamefont
  {K.}~\bibnamefont {Saito}},\ and\ \bibinfo {author} {\bibfnamefont
  {K.}~\bibnamefont {Kohn}},\ }\bibfield  {title} {\bibinfo {title} {Magnetic
  ordering in one-dimensional system {C}o{N}b$_2${O}$_6$ with competing
  interchain interactions},\ }\href@noop {} {\bibfield  {journal} {\bibinfo
  {journal} {J. Phys. Soc. Jpn.}\ }\textbf {\bibinfo {volume} {63}},\ \bibinfo
  {pages} {3568} (\bibinfo {year} {1994})}\BibitemShut {NoStop}%
\bibitem [{\citenamefont {Heid}\ \emph {et~al.}(1995)\citenamefont {Heid},
  \citenamefont {Weitzel}, \citenamefont {Burlet}, \citenamefont {Bonnet},
  \citenamefont {Gonschorek}, \citenamefont {Vogt}, \citenamefont {Norwig},\
  and\ \citenamefont {Fuess}}]{Heid1995bt}%
  \BibitemOpen
  \bibfield  {author} {\bibinfo {author} {\bibfnamefont {C.}~\bibnamefont
  {Heid}}, \bibinfo {author} {\bibfnamefont {H.}~\bibnamefont {Weitzel}},
  \bibinfo {author} {\bibfnamefont {P.}~\bibnamefont {Burlet}}, \bibinfo
  {author} {\bibfnamefont {M.}~\bibnamefont {Bonnet}}, \bibinfo {author}
  {\bibfnamefont {W.}~\bibnamefont {Gonschorek}}, \bibinfo {author}
  {\bibfnamefont {T.}~\bibnamefont {Vogt}}, \bibinfo {author} {\bibfnamefont
  {J.}~\bibnamefont {Norwig}},\ and\ \bibinfo {author} {\bibfnamefont
  {H.}~\bibnamefont {Fuess}},\ }\bibfield  {title} {\bibinfo {title} {Magnetic
  phase-diagram of {C}o{N}b$_2${O}$_6$ - a neutron-diffraction study},\
  }\href@noop {} {\bibfield  {journal} {\bibinfo  {journal} {J. Magn. Magn.
  Mater.}\ }\textbf {\bibinfo {volume} {151}},\ \bibinfo {pages} {123}
  (\bibinfo {year} {1995})}\BibitemShut {NoStop}%
\bibitem [{\citenamefont {Weitzel}\ \emph {et~al.}(2000)\citenamefont
  {Weitzel}, \citenamefont {Ehrenberg}, \citenamefont {Heid}, \citenamefont
  {Fuess},\ and\ \citenamefont {Burlet}}]{Weitzel2000ei}%
  \BibitemOpen
  \bibfield  {author} {\bibinfo {author} {\bibfnamefont {H.}~\bibnamefont
  {Weitzel}}, \bibinfo {author} {\bibfnamefont {H.}~\bibnamefont {Ehrenberg}},
  \bibinfo {author} {\bibfnamefont {C.}~\bibnamefont {Heid}}, \bibinfo {author}
  {\bibfnamefont {H.}~\bibnamefont {Fuess}},\ and\ \bibinfo {author}
  {\bibfnamefont {P.}~\bibnamefont {Burlet}},\ }\bibfield  {title} {\bibinfo
  {title} {Lifshitz point in the three-dimensional magnetic phase diagram of
  {C}o{N}b$_2${O}$_6$},\ }\href@noop {} {\bibfield  {journal} {\bibinfo
  {journal} {Phys. Rev. B}\ }\textbf {\bibinfo {volume} {62}},\ \bibinfo
  {pages} {12146} (\bibinfo {year} {2000})}\BibitemShut {NoStop}%
\bibitem [{\citenamefont {Fava}\ \emph {et~al.}(2020)\citenamefont {Fava},
  \citenamefont {Coldea},\ and\ \citenamefont {Parameswaran}}]{Fava2020}%
  \BibitemOpen
  \bibfield  {author} {\bibinfo {author} {\bibfnamefont {M.}~\bibnamefont
  {Fava}}, \bibinfo {author} {\bibfnamefont {R.}~\bibnamefont {Coldea}},\ and\
  \bibinfo {author} {\bibfnamefont {S.~A.}\ \bibnamefont {Parameswaran}},\
  }\bibfield  {title} {\bibinfo {title} {Glide symmetry breaking and {I}sing
  criticality in the quasi-1d magnet {C}o{N}b$_2${O}$_6$},\ }\href
  {https://doi.org/10.1073/pnas.2007986117} {\bibfield  {journal} {\bibinfo
  {journal} {Proc. Natl. Acad. Sci. USA}\ }\textbf {\bibinfo {volume} {117}},\
  \bibinfo {pages} {25219} (\bibinfo {year} {2020})}\BibitemShut {NoStop}%
\bibitem [{\citenamefont {Morris}\ \emph {et~al.}(2021)\citenamefont {Morris},
  \citenamefont {Desai}, \citenamefont {Viirok}, \citenamefont {H{\"u}vonen},
  \citenamefont {Nagel}, \citenamefont {R{\~o}{\~o}m}, \citenamefont {Krizan},
  \citenamefont {Cava}, \citenamefont {McQueen}, \citenamefont {Koohpayeh},
  \citenamefont {Kaul},\ and\ \citenamefont {Armitage}}]{Morris2021}%
  \BibitemOpen
  \bibfield  {author} {\bibinfo {author} {\bibfnamefont {C.~M.}\ \bibnamefont
  {Morris}}, \bibinfo {author} {\bibfnamefont {N.}~\bibnamefont {Desai}},
  \bibinfo {author} {\bibfnamefont {J.}~\bibnamefont {Viirok}}, \bibinfo
  {author} {\bibfnamefont {D.}~\bibnamefont {H{\"u}vonen}}, \bibinfo {author}
  {\bibfnamefont {U.}~\bibnamefont {Nagel}}, \bibinfo {author} {\bibfnamefont
  {T.}~\bibnamefont {R{\~o}{\~o}m}}, \bibinfo {author} {\bibfnamefont {J.~W.}\
  \bibnamefont {Krizan}}, \bibinfo {author} {\bibfnamefont {R.~J.}\
  \bibnamefont {Cava}}, \bibinfo {author} {\bibfnamefont {T.~M.}\ \bibnamefont
  {McQueen}}, \bibinfo {author} {\bibfnamefont {S.~M.}\ \bibnamefont
  {Koohpayeh}}, \bibinfo {author} {\bibfnamefont {R.~K.}\ \bibnamefont
  {Kaul}},\ and\ \bibinfo {author} {\bibfnamefont {N.~P.}\ \bibnamefont
  {Armitage}},\ }\bibfield  {title} {\bibinfo {title} {Duality and domain wall
  dynamics in a twisted {K}itaev chain},\ }\href
  {https://doi.org/10.1038/s41567-021-01208-0} {\bibfield  {journal} {\bibinfo
  {journal} {Nat. Phys.}\ }\textbf {\bibinfo {volume} {17}},\ \bibinfo {pages}
  {832} (\bibinfo {year} {2021})}\BibitemShut {NoStop}%
\bibitem [{\citenamefont {Woodland}\ \emph {et~al.}(2023)\citenamefont
  {Woodland}, \citenamefont {Macdougal}, \citenamefont {Cabrera}, \citenamefont
  {Thompson}, \citenamefont {Prabhakaran}, \citenamefont {Bewley},\ and\
  \citenamefont {Coldea}}]{Woodland2023}%
  \BibitemOpen
  \bibfield  {author} {\bibinfo {author} {\bibfnamefont {L.}~\bibnamefont
  {Woodland}}, \bibinfo {author} {\bibfnamefont {D.}~\bibnamefont {Macdougal}},
  \bibinfo {author} {\bibfnamefont {I.~M.}\ \bibnamefont {Cabrera}}, \bibinfo
  {author} {\bibfnamefont {J.~D.}\ \bibnamefont {Thompson}}, \bibinfo {author}
  {\bibfnamefont {D.}~\bibnamefont {Prabhakaran}}, \bibinfo {author}
  {\bibfnamefont {R.~I.}\ \bibnamefont {Bewley}},\ and\ \bibinfo {author}
  {\bibfnamefont {R.}~\bibnamefont {Coldea}},\ }\bibfield  {title} {\bibinfo
  {title} {Tuning the confinement potential between spinons in the {I}sing
  chain compound {C}o{N}b$_2${O}$_6$ using longitudinal fields and quantitative
  determination of the microscopic {H}amiltonian},\ }\href
  {https://doi.org/10.1103/PhysRevB.108.184416} {\bibfield  {journal} {\bibinfo
   {journal} {Phys. Rev. B}\ }\textbf {\bibinfo {volume} {108}},\ \bibinfo
  {pages} {184416} (\bibinfo {year} {2023})}\BibitemShut {NoStop}%
\bibitem [{\citenamefont {Prabhakaran}\ \emph {et~al.}(2003)\citenamefont
  {Prabhakaran}, \citenamefont {Wondre},\ and\ \citenamefont {Boothroyd}}]{PB}%
  \BibitemOpen
  \bibfield  {author} {\bibinfo {author} {\bibfnamefont {D.}~\bibnamefont
  {Prabhakaran}}, \bibinfo {author} {\bibfnamefont {F.~R.}~\bibnamefont
  {Wondre}},\ and\ \bibinfo {author} {\bibfnamefont {A.~T.}~\bibnamefont
  {Boothroyd}},\ }\bibfield  {title} {\bibinfo {title} {Preparation of large
  single crystals of {A}{N}b$_2${O}$_6$ ({A}={N}i, {C}o, {F}e, {M}n) by the
  floating-zone method},\ }\href
  {https://doi.org/https://doi.org/10.1016/S0022-0248(02)02229-7} {\bibfield
  {journal} {\bibinfo  {journal} {J. Cryst. Growth}\ }\textbf {\bibinfo
  {volume} {250}},\ \bibinfo {pages} {72} (\bibinfo {year} {2003})}\BibitemShut
  {NoStop}%
\bibitem [{\citenamefont {Wheeler}(2007)}]{Wheeler2007}%
  \BibitemOpen
  \bibfield  {author} {\bibinfo {author} {\bibfnamefont {E.~M.}\ \bibnamefont
  {Wheeler}},\ }\emph {\bibinfo {title} {Neutron scattering from
  low-dimensional quantum magnets}},\ \href@noop {} {Ph.D. thesis},\ \bibinfo
  {school} {University of Oxford} (\bibinfo {year} {2007})\BibitemShut
  {NoStop}%
\bibitem [{dat()}]{database}%
  \BibitemOpen
  \href {http://dx.doi.org/10.5287/ora-braxw4ez1} {\bibinfo {title} {Data
  archive weblink}},\ \bibinfo {howpublished}
  {\url{http://dx.doi.org/10.5287/ora-braxw4ez1}}\BibitemShut {NoStop}%
\end{thebibliography}
\end{document}